# The leaky integrator that could:

# Or recursive polynomial regression for online signal analysis

Hugh L Kennedy

*DEWC Services (hugh.kennedy@dewc.com)*

*UniSA STEM Unit, University of South Australia (hugh.kennedy@unisa.edu.au)*

*Mawson Lakes SA 5095*

## Abstract

Fitting a local polynomial model to a noisy sequence of uniformly sampled observations or measurements (i.e. regressing) by minimizing the sum of weighted squared errors (i.e. residuals) may be used to design digital filters for a diverse range of signal-analysis problems, such as detection, classification and tracking (i.e. smoothing or state estimation), in biomedical, financial, and aerospace applications, for instance. Furthermore, the recursive realization of such filters, using a network of so-called 'leaky' integrators, yields simple digital components with a low computational complexity and an infinite impulse response (IIR) that are ideal in embedded online sensing systems with high data rates. The variance of the smoothed output or the covariance matrix for the vector of derivative states are also produced. These quantities may be used to support constant false-alarm rates in detection systems or to define adaptive gates in target trackers. Target tracking, pulse-edge detection, peak detection and anomaly/change detection are considered in this tutorial as illustrative examples.

Erlang-weighted polynomial regression provides a design framework within which the various design trade-offs of state estimators (e.g. bias errors vs. random errors) and IIR smoothers (e.g. frequency isolation vs. time localization or linearity vs. latency) may be intuitively balanced. Erlang weights are configured using a smoothing parameter which determines the decay rate of the exponential 'tail'; and a shape parameter which may be used to discount more recent data, so that a greater relative emphasis is placed on a past time interval. In Morrison's 1969 treatise on sequential smoothing and prediction, the exponential weight (i.e. the zero shape-parameter case, for a fading memory) and the Laguerre polynomials that are orthogonal with respect to this weight, are described in detail; however, more general Erlang weights (with non-negative shape parameters for lagged fading memories) and the resulting *associated* Laguerre polynomials are not considered there, nor have they been covered in detail elsewhere since. Thus, one of the purposes of this tutorial is to explain how Erlang weights may be used to shape and improve the (impulse and frequency) response of recursive regression filters.

# Contents

# Definitions

$p$: Smoothing parameter on the real $(0,1)$ interval in the complex $z$-plane. Determines the rate of exponential decay of the Erlang regression weight. A dimensionless quantity.

$n$: Sample index, in discrete time, for $n = 0 \ldots N - 1$.

$N$: Number of samples.

$x[n]$: Digital filter input at the time of the $n$th sample.

$y[n]$: Digital filter output at the time of the $n$th sample.

[■]: Used to index (from zero) an infinite sequence of samples or the elements of a finite vector.

$z^{-1}$: Unit delay operator. The $\mathcal{Z}$-transform of a digital component that delays its input by one sample.

$z$: The complex $z$-plane coordinate. The $\mathcal{Z}$-transform of a unit advance operation. Is non-causal and un-realizable but positive powers of $z$ are more convenient for system analysis.

$e^{-i\omega}$: The frequency response of a unit delay operation.

$i$: The complex unit.

$\omega$: Relative angular frequency (radians per sample).

$f$: Relative frequency (cycles per sample) $f = \omega/2\pi$.

$m$: Delay index, for $m = 0 \ldots M - 1$ (finite weight) or $m = 0 \ldots \infty$ (infinite weight); $m = 0$ corresponds to the current (i.e. most recent) sample.

$M$: The length (in samples) of a finite weight.

$w[m]$: The weight applied to the residuals for the least-squares fit. Used to emphasise more relevant samples times in the regression operation.

$\lambda_w$: Time-constant/scale (in samples) of the exponential decay of the regression weight.

$\mathcal{H}_{y \leftarrow x}(z)$: Discrete-time transfer function of a digital filter. Connects the input $x[n]$ to the output $y[n]$.

$\mathcal{X}(z)$: $\mathcal{Z}$-transform of filter input $x[n]$.

$\mathcal{Y}(z)$: $\mathcal{Z}$-transform of filter input $y[n]$.

$\kappa$: Shape parameter of an Erlang weight. A non-negative integer, i.e. $\kappa \geq 0$.

$\mu_w$: Mean of continuous Erlang weight.

$\sigma_w^2$: Variance of continuous Erlang weight.

$\mu_w$: Skew of continuous Erlang weight.

$w_k[n]$ The $k$th element of the internal state of a linear state-space (LSS) system in discrete time at the time of the $n$th sample, for $k = 0 \ldots K - 1$.

$\boldsymbol{w}[n]$: A $K \times 1$ vector that holds the $K$ internal states of a discrete-time LSS system at the time of the $n$th sample.

$K$: The order of a discrete-time LSS system.

$\boldsymbol{G}$: The $K \times K$ state propagation matrix of a discrete-time LSS system.

$\boldsymbol{H}$: The $K \times 1$ input vector of a discrete-time LSS system.

$\boldsymbol{C}$: The $1 \times K$ output vector of a discrete-time LSS system.

$\phi_k[m]$: Impulse response of the discrete-time transfer-function that links $x[n]$ to $w_k[n]$.

$\boldsymbol{\gamma}$: A $K \times K$ matrix of zeros with normalizing factors $\gamma_k$ along the diagonal.

$\gamma_k$: Normalizing factor of an Erlang weight with $\kappa = k$.

$\boldsymbol{T}_{w \leftarrow \phi}$: A $K \times K$ lower triangle matrix (i.e. with zeros above the diagonal) that transforms state impulse responses $\phi$ into the Erlang weights $w$.

$S_k(p)$: Infinite summation of a discrete-time Erlang weight with $\kappa = k$.

$\boldsymbol{\rho}$: A $K \times 1$ vector of constants that are set using the final-value theorem. Used to initialize the internal states of a discrete-time LSS, to reduce the start-up transient.

$\rho_k$: The $k$th element of $\boldsymbol{\rho}$, for $k = 0 \ldots K - 1$.

$\mathcal{H}_{w_k \leftarrow x}(z)$: The discrete-time transfer function that links $x[n]$ to $w_k[n]$.

$X$: An unknown parameter to be estimated. A constant.

$Y[n]$: The $n$th measurement or observation of $X$. A random variable.

$\varepsilon[n]$: The measurement error added to $X$ to form $Y[n]$.

$\sigma_\varepsilon^2$: The variance of the measurement error.

$E\langle\blacksquare\rangle$: The expectation of a random variable.

$\mathrm{Cov}\langle\blacksquare,\blacksquare\rangle$: The covariance of two random variables.

$K_X$: The order of the polynomial regression model. The degree of the polynomial model (i.e. the highest power) is equal to $K_X - 1$.

$\alpha_{k_X}$: The $k_X$th polynomial model coefficient, i.e. for the $m^{k_X}$ monomial term, for $k_X = 0 \ldots K_X - 1$.

$\varphi_{k_X}[m]$: The $k_X$th monomial component or term, i.e. $\varphi_{k_X}[m] = m^{k_X}$.

$S_\varepsilon$: The weighted sum-of-squared residuals for the polynomial fit to the measurements.

$\boldsymbol{\alpha}$: The $K_X \times 1$ coefficient vector with elements $\hat{\alpha}_{k_X}$.

$\hat{\alpha}_{k_X}$: The estimate of the $\alpha_{k_X}$ model parameter.

$\boldsymbol{y}$: The $M \times 1$ data vector with the $m$th element equal to $Y[n-m]$.

$\boldsymbol{X}_\varphi$: The $M \times K_X$ regressor matrix with elements $\varphi_{k_X}[m] = m^{k_X}$.

$\boldsymbol{\mathcal{W}}$: The $M \times M$ 'weight' matrix of zeros with $\mathcal{W}[m,m] = w[m]$ along the diagonal.

$[\blacksquare, \blacksquare]$: Used to index (from zero) the elements of a matrix.

$k_a$: A row index.

$k_b$: A column index.

$\boldsymbol{x}$: The $M \times 1$ vector with the $m$th element equal to $X[m]$.

$\blacksquare^T$: The matrix/vector transpose operator.

$\blacksquare^{-1}$: The matrix inverse operator.

$\boldsymbol{S}_{\varphi \mathcal{W} \varphi}$: The $K_X \times K_X$ weighted overlap matrix (between the monomial components).

$\boldsymbol{s}_{\varphi \mathcal{W} Y}$: The $K_X \times 1$ weighted projection vector (of the data onto the monomial components).

$s_{Y \mathcal{W} Y}$: The weighted sum of squared measurement inputs.

$\hat{\sigma}_\varepsilon^2$: The estimate of the measurement error variance, derived from the residual of the weighted least-squares fit.

$\psi_{k_X}[m]$: The $k_X$th orthonormal component, with respect to the weight. These components are the discrete-time Laguerre polynomials for $\kappa = 0$ or the discrete-time associated Laguerre polynomials for $\kappa > 0$.

$\beta_{k_X}$: The model coefficient of the $k_X$th orthonormal component.

$\boldsymbol{S}_{\psi \mathcal{W} \psi}$: The $K_X \times K_X$ weighted overlap matrix (between the orthonormal components). Is equal to the identity matrix.

$\boldsymbol{s}_{\psi \mathcal{W} Y}$: The $K_X \times 1$ weighted projection vector (of the data onto the orthonormal components).

$\boldsymbol{I}_{K \times K}$: The $K \times K$ identity matrix, i.e. with ones along the diagonal and zeros elsewhere.

$\boldsymbol{0}_{K \times K}$: A $K \times K$ matrix of zeros.

$\boldsymbol{X}_\psi$: The $M \times K_X$ regressor matrix with elements $\psi_{k_X}[m]$.

$\boldsymbol{T}_{\psi \leftarrow \varphi}$: The $K_X \times K_X$ (lower-triangle) matrix that transforms the monomial model components into the orthonormal model components.

$\boldsymbol{T}_{\varphi \leftarrow \psi}$: The $K_X \times K_X$ (upper-triangle) matrix that transforms the orthonormal model components into the monomial model components.

$\boldsymbol{\beta}$: The $K_X \times 1$ coefficient vector with elements $\hat{\beta}_{k_X}$.

$\hat{\beta}_{k_X}$: The estimate of the $\beta_{k_X}$ model parameter.

$\hat{X}$: The weighted least-squares estimate of $X$ at a given point in (continuous) time.

$\boldsymbol{y}_1, \boldsymbol{C}_1, \boldsymbol{G}_1, \boldsymbol{H}_1, \boldsymbol{x}_1, \boldsymbol{w}_1, K_1$: The discrete-time LSS system used to recursively compute Erlang-weighted means of the input, for a series of $K_1$ Erlang weights with $\kappa = 0 \ldots K_1 - 1$. This 'first-moment' recursion is used to estimate the model parameters and the measurement error variance. Note that $K_1 = \kappa + K_X$ and $\boldsymbol{C}_1$ is a $K_X \times K_1$ matrix.

$\boldsymbol{y}_2, \boldsymbol{C}_2, \boldsymbol{G}_2, \boldsymbol{H}_2, \boldsymbol{x}_2, \boldsymbol{w}_2, K_2$: The discrete-time LSS system used to recursively compute weighted mean of the squared input, for an Erlang weight with a shape parameter of $\kappa$. This 'second-moment' recursion is used to estimate the measurement error variance and the covariance matrix of the model parameters or state estimates. Note that $K_2 = \kappa$ and $\boldsymbol{C}_2$ is a $1 \times K_2$ vector.

$\sigma_0^2$: A coarse estimate of the measurement error variance, i.e. $\sigma_\varepsilon^2$. Used to initialize the second-moment recursion.

$T_s$: The sampling period (seconds). Assumed to be constant.

$t$: The continuous time variable (seconds), $t = T_s n$.

$\tau$: A time offset (seconds), relative to the current time, $\tau = -T_s m$.

$X(\tau)$: The expected value of $Y(t + \tau)$.

$X^{(k_t)}(\tau)$: The $k_t$th derivative of $X(\tau)$ with respect to time, for $k_t = 0 \dots K_t - 1$.

$K_t$: The number of temporal derivatives to estimate.

$\hat{X}^{(k_t)}(\tau)$: The weighted least-squares estimate of $X^{(k_t)}(\tau)$.

$\boldsymbol{y}_q[n]$: A $K_t \times 1$ vector of derivative state estimates with the $k_t$th element equal to $\hat{X}^{(k_t)}(\tau)$ with $\tau = -T_s q$.

$q$: The delay parameter. A relative delay with a non-integer value (in samples).

$\boldsymbol{C}_q$: The $K_t \times K_1$ output matrix of the recursive estimator of derivative states.

$\boldsymbol{D}_q$: The $K_t \times K_X$ synthesis matrix, that differentiates and evaluates the fitted polynomial model at time $t = T_s(n - q)$.

$\hat{\sigma}_{k_t}^2$: the estimated variance of the $k_t$th estimate, i.e. $X^{(k_t)}(\tau)$.

$\boldsymbol{\Sigma}$: The $K_t \times K_t$ covariance matrix of the vector containing the derivative state estimates.

VRF: Variance reduction factor.

WNG: White-noise gain.

$h_{k_t}[m]$: The impulse response of the discrete-time transfer function that links the input $x[n]$ to the $k_t$th element of the output vector $\boldsymbol{y}_q[n]$.

$H_{k_t}(\omega)$: The frequency response of the discrete-time transfer function that links the input $x[n]$ to the $k_t$th element of the output vector $\boldsymbol{y}_q[n]$.

**VRF**: A $K_t \times K_t$ matrix of variance reduction factors. Used to construct the $\boldsymbol{\Sigma}$ matrix.

$\boldsymbol{C}_\varphi$: A $K_t \times K_X$ matrix. Used to construct the **VRF** matrix.

$\mathcal{V}_{k_t}(q)$: The variance reduction factor of the $k_t$th estimate, as a function of the delay parameter.

$\dot{\mathcal{V}}_{k_t}(q)$. The first derivative of $\mathcal{V}_{k_t}(q)$ with respect to $q$.

$1/s$: The Laplace transform of an integration operation in continuous time.

$s$: The complex $s$-plane Coordinate. The Laplace transform of an (ideal) differentiation operation in continuous time.

$\Delta_t$: Uncertainty in (continuous) time.

$\Delta_\Omega$: Uncertainty in frequency.

$\Omega$: Angular frequency (radians per second). $\Omega = \omega/T_s$.

$\sigma_x$: Standard deviation of position.

$\sigma_v$: Standard deviation of momentum.

$\hbar$: Plank's constant.

$\sigma_t^2$: Variance in time (seconds squared).

$\sigma_\Omega^2$: Variance in angular frequency (radians squared over seconds squared).

$m_t^k$: The $k$th moment of a pulse in time.

$m_\Omega^k$: The $k$th moment of a pulse in frequency.

$\psi(t)$: The pulse magnitude (a non-negative function) in the time domain.

$\Psi(\Omega)$: The pulse magnitude (a non-negative function) in the frequency domain.

$\mathcal{F}\{\blacksquare\}$: The Fourier-transform operator.

$\lambda_t$: The scale (in seconds) of an Erlang pulse in time.

$\mathcal{L}\{\blacksquare\}$: The Laplace-transform operator.

$\Psi(s)$: The Laplace transform of $\psi(t)$.

$\Gamma(\blacksquare)$: The Gamma function.

$D(\omega)$: The desired low-frequency response of a smoother.

$H(\omega)$: The actual frequency response of a smoother.

$E(\omega)$: The frequency response error of a smoother.

$\omega_c$: The distortion-free bandwidth (radians per sample) of a smoother.

$f_c$: The distortion-free bandwidth (cycles per sample) of a smoother.

$|\blacksquare|$: The magnitude of the complex argument.

$\angle\blacksquare$: The angle (radians) of the complex argument.

$k_\omega$: The derivative order, of a complex frequency response, with respect to $\omega$.

$Z$: The test statistic considered by a detector.

$\lambda_Z$: The detection threshold applied by a detector.

$\hat{A}$ & $\hat{B}$: The mean outputs of filters A & B.

$\hat{a}^2$ & $\hat{b}^2$: The variance outputs of filters A & B.

$q_A$ & $q_B$. The (low-frequency) group-delay applied by filters A & B.

$\text{VRF}_A$ & $\text{VRF}_B$. The variance reduction factors of the smoothers in filters A & B.

# Introduction

As succinctly stated in Morrison's 1969 treatise [1]: "by virtue of their extreme simplicity, polynomials give rise to extremely compact smoothing algorithms. Moreover they can generally be used for smoothing over short enough intervals with very little actual knowledge of the true process". Regression of a polynomial model is one of the most basic forms of data analysis and its efficient recursive realization allows standard statistical methods to be applied in embedded online systems with high data rates [2]. Recursion allows the parameters derived from long data records to be sequentially updated by simply operating on the new input sample (i.e. an observation or measurement) and the old internal state of the estimator.

Expositions on the theory and application of digital filters usually begin with the mathematical formalities then move towards the details of their computer realization. In the approach taken here, we proceed in the reverse direction: The fundamental statistical operation of analysis, i.e. the moving average, and the atomic digital unit of its implementation, i.e. the so-called 'leaky' integrator, are introduced at the outset. More complicated operations and their corresponding digital realizations are then developed from these basic concepts/components and their responses analysed then applied in the problems mentioned above. It is hoped that this 'inside-out' approach will appeal to software/firmware/hardware practitioners with an interest in implementation.

The coefficients of these recursive regression filters with an infinite impulse response (IIR) are derived in the time domain [3]; therefore, some of the more abstract concepts of signals-and-systems theory [4][5], that are usually required to design low-pass and band-pass IIR filters, such as pole/zero representations in the complex $z$-plane, may be (largely) avoided [6][7]. Furthermore, these filters do not attempt to leverage prior knowledge of (process and measurement) noise distributions, thus an appreciation of Bayesian inference, and the availability of noise parameters, is not required [8][9]. Linear least-squares regression of a polynomial model is one of the most basic (and oldest) forms of data analysis [10]. It is hoped that this way of designing IIR filters and recursive state estimators will appeal to readers with an interest in application who are unfamiliar with optimal filtering theories.

In Section 1 the leaky integrator is introduced as a way of computing an exponential moving average; then a network of leaky integrators is used to recursively compute an Erlang-weighted moving average. The network is then realized via a discrete-time linear state-space model to implement weighted polynomial regression in Section 2 and used to estimate derivative states and the corresponding covariance matrix in Section 3. The way in which the complex frequency response (i.e. magnitude and phase) of the regression filter (e.g. a smoother or differentiator) is shaped by the form of the regression weight, the order of the polynomial model, and the point in time at which the polynomial or its derivatives are evaluated, is then discussed in Section 4. Target tracking, pulse-edge detection, peak detection and change/anomaly detection are considered in Section 5 as illustrative applications.

## Section 1. Recursive moving averages

Digital integrators (or accumulators) are the fundamental atomic elements of recursive regression algorithms. A *lossless* integrator simply outputs the sum of all prior inputs, and such units are ideal for the evaluation of very long-term averages for the removal of random errors from observations of static systems. In a *leaky* integrator, the internal sum in computer memory decays at a pre-determined rate. The sum is smaller by a factor of $p$ for each new zero input, where $p$ is a 'smoothing' parameter on the $(0,1)$ interval; thus, the influence of older inputs on the sum is diluted (with exponential decay) over time. Older inputs are 'forgotten' rapidly as $p \to 0$ (from the right) and slowly as $p \to 1$ (from the left), for high-gain (i.e. less smoothing) and low-gain (i.e. more smoothing) filters, respectively. The simplicity of these digital components is apparent in Figure 1.

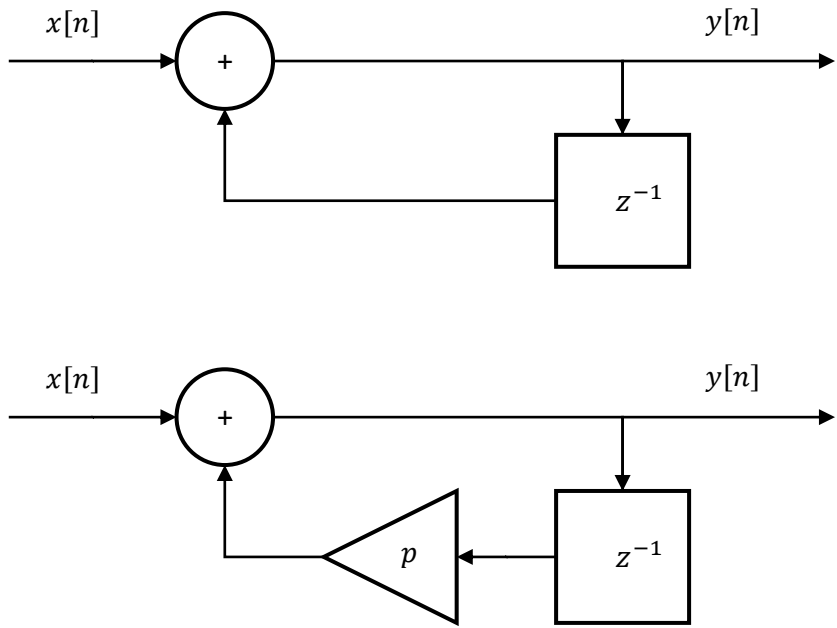

*Figure 1. Digital integrator circuits. Lossless integrator (top); leaky integrator (bottom). In these circuits, the $z^{-m}$ block is an m-sample delay operator, and p is a scaling factor that is greater than zero and less than one.*

The *fading* memory (i.e. wider bandwidth) of the leaky integrator is used in filters that estimate the states of dynamic systems that are routinely perturbed by random inputs (e.g. operator commands or environmental disturbances). The ever-*expanding* memory of the (narrow band) lossless integrator may be used to estimate the state of ideal unperturbed systems thus it is of little use in practice. As discussed in more detail below, sums with a *finite* memory are formed by delaying and subtracting the output of a lossless integrator.

Moving averages are the most basic form of online statistical analysis (a regression model with a single constant term) and they are often used to remove noise from time-series data that are sampled at a uniform rate.

In a *simple* moving average (with a finite memory) the same weight is applied to all samples within a sliding 'window', so they influence the computed mean (and optionally, the variance) equally [13]. Furthermore, for an $M$-point simple moving average, only the $M$ most recent samples are considered, all older samples are ignored, and ignorance of all future samples is assumed. Thus, this average uses a uniform weight of finite extent, or a 'rectangular' weight, i.e.

$w[m] = 1/M$ (for $m = 0 \dots M - 1$) and  (1a)

$w[m] = 0$ (for $m = M \dots \infty$), where  (1b)

$m$ is the delay index, into past ($m > 0$) and present ($m = 0$) samples.  (1c)

Other windows of finite extent but with a non-constant weight (e.g. a truncated Gaussian or a Slepian) could also be used for moving average filters with a finite impulse response (FIR); however, these (endless) possibilities are not considered further here.

As the name suggests, *exponential* moving averages (with a fading memory) apply an exponentially decaying weight so that older samples are de-emphasised but never forgotten completely [13]. This average uses a weight of

$w[m] = (1-p)p^m$ (for $m = 0 \dots \infty$) with (2a)

$p = e^{-1/\lambda_w}$ where (2b)

$\lambda_w$ is the time-constant/scale or 'centroid' of the exponential (in samples) such that (2c)

$\sum_{m=0}^{\infty} m w[m] = \lambda_w$ (with $\lambda_w > 1$) and where the factor of (2d)

$(1-p)$ is a normalizing term which ensures that (2e)

$\sum_{m=0}^{\infty} w[m] = 1$. (2f)

Moving averages are computed non-recursively using

$y[n] = \sum_{m=0}^{\infty} w[m] x[n-m]$, where (3a)

$w[m]$ is the weight, (3b)

$x[n]$ is the $n$th sample (or input) with $x[n] = 0$ for $n < 0$ and (3c)

$y[n]$ is the computed mean (or output) at the time of the $n$th sample (for $n = 0 \dots \infty$). (3d)

For a rectangular weight, the infinite sum is truncated early, after $m = M - 1$. For an exponential weight, the infinite sum may be truncated after it converges to within a specified tolerance. Both weights may also be applied recursively using digital integrators.

For a recursive simple moving average, the output of a lossless integrator is delayed by $M$ samples and subtracted from the current output to form the finite summation over the $M$-sample rectangular window (see Figure 2). However, for inputs that are finite, constant and non-zero, the output of the lossless integrator grows without bound, thus the recursion is susceptible to the accumulation of rounding errors, when very large numbers are differenced, on finite-precision machines.

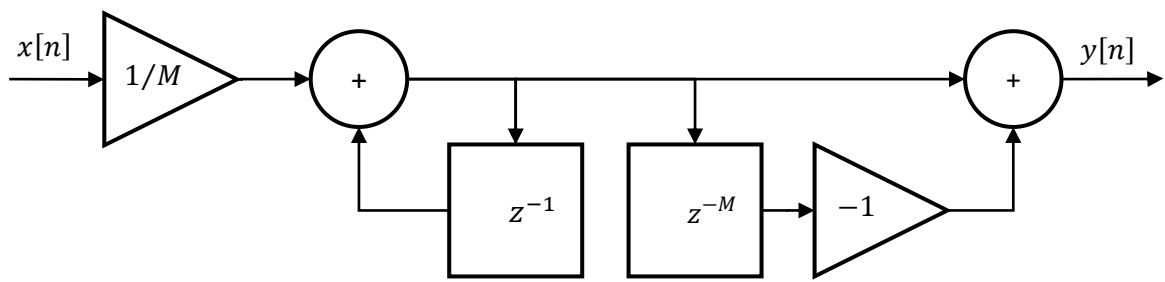

*Figure 2. A digital circuit for the recursive realization of a simple M-sample moving average, featuring from left to right: a normalizing factor, an ideal integrator (with feedback) and a delayed subtraction (with 'feedforward') i.e. a so-called 'comb'.*

The exponential moving average also involves an infinite summation; however, the summation decays exponentially over time – rapidly as $p \to 0$ (from the right) and slowly as $p \to 1$ (from the left) – so it does not grow without bound. It is computed recursively and robustly using a leaky integrator in a digital circuit (see Figure 3) or in one line of computer code using

$y[n] = p y[n-1] + (1-p) x[n]$. (4a)

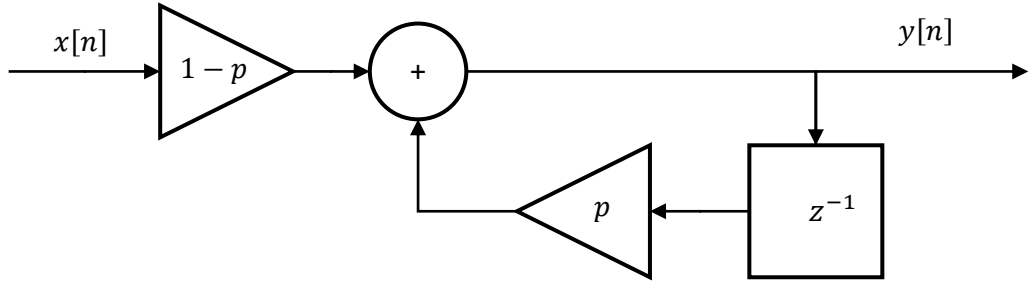

*Figure 3. A digital circuit for the recursive realization of an exponential moving average, featuring from left to right: a normalizing factor and a leaky integrator.*

The discrete-time transfer function of this system is $\mathcal{H}_{y \leftarrow x}(z)$; it leads directly to the frequency response of the system although it is not essential for the derivation or the implementation of the filters described here [6]. This function connects the input $x[n]$ to the output $y[n]$ in the discrete-time domain and $\mathcal{X}(z)$ to $\mathcal{Y}(z)$ in complex $z$-domain, i.e.

$$\mathcal{Y}(z) = \mathcal{H}_{y \leftarrow x}(z)\mathcal{X}(z) \text{ or} \tag{5a}$$

$$\mathcal{H}_{y \leftarrow x}(z) = \mathcal{Y}(z)/\mathcal{X}(z), \text{ by definition, where} \tag{5b}$$

$$\mathcal{X}(z) \text{ \& } \mathcal{Y}(z) \text{ are the transforms of } x[n] \text{ \& } y[n]. \tag{5c}$$

For the circuit in Figure 3:

$$\mathcal{H}_{y \leftarrow x}(z) = (1-p)z/(z-p). \tag{6a}$$

It is reached by taking $\mathcal{Z}$ transforms of all terms in (4a):

$$\mathcal{Y}(z) = pz^{-1}\mathcal{Y}(z) + (1-p)\mathcal{X}(z) \text{ and rearranging} \tag{7a}$$

$$\mathcal{Y}(z) - pz^{-1}\mathcal{Y}(z) = (1-p)\mathcal{X}(z) \tag{7b}$$

$$\mathcal{Y}(z)(1-pz^{-1}) = (1-p)\mathcal{X}(z) \text{ thus} \tag{7c}$$

$$\mathcal{Y}(z)/\mathcal{X}(z) = (1-p)/(1-pz^{-1}) \text{ and} \tag{7d}$$

$$\mathcal{H}_{y \leftarrow x}(z) = (1-p)z/(z-p). \tag{7e}$$

The rectangular weight used in a simple moving average is symmetric about its 'centre-of-mass' or centroid, at $m = (M-1)/2$, but truncates abruptly and becomes zero at both ends of the time 'window' or time 'horizon', i.e. at $m < 0$ due to the unavailability of future data and at $m > M - 1$ due to the irrelevance of older data. In the time domain, this yields a sudden step up and down when a large input enters and leaves the sliding window, respectively. In the frequency domain, the perfect symmetry of the rectangular weight yields a *linear phase response*, and the truncation establishes anti-resonant modes which prevent sinusoidal components from affecting the computed average when the length of the window is a whole multiple of the sinusoid's wavelength, i.e. at nulls of the frequency response. Sinusoids with incommensurate wavelengths do make a small contribution to the mean, 'via the sidelobes' of the window's frequency response.

The weight used in an exponential moving average decays gradually as $m$ increases, which is sufficient to eliminate steps in time caused by 'aged-out' observations and nulls/sidelobes in frequency caused by anti-resonance. However, it does still feature abrupt truncation at $m = 0$ (because the future is unknown) thus entry steps remain. Moreover, the weight now has no centre of symmetry in the time domain, thus the exponential moving average has a *non-linear phase-response*.

When the moving average of a sinusoidal input is computed, the output of the filter is a scaled and delayed sinusoid of the same frequency. The weights used in simple and exponential averages have a low-pass response therefore high frequencies experience magnitude attenuation and all frequencies experience a phase lag. The *linear* phase-response of the *simple* moving average means that the delay is equal to $(M-1)/2$, regardless of frequency. However, the *non-linear* phase-response of the *exponential* moving average means that the delay is frequency dependent; although, the dispersion is negligible for very low frequencies that are well within the 'passband' of the moving-average filter. The step of the exponential weight at $m = 0$ may be 'softened', and the asymmetry (i.e. its skew) about its centroid (i.e. its mean) may be decreased, by cascading multiple exponentials in series, for more frequency-selective Erlang weights that have a passband with a flatter magnitude-response, a *more-linear* phase-response and improved high-frequency noise attenuation, at the expense of an increased average passband delay. The response characteristics of the various weights are examined more closely in Section 3.

Phase lags should be minimised for robust stability in feedback control loops (e.g. guided-missile seekers) and for timely alerts from automated decision systems (e.g. early-warning and self-protection). However, in some online systems, where rapid response times are not critical, longer lags are acceptable if they improve accuracy and error rates by reducing noise (in the high-frequency 'stopband') and signal distortion (in the low-frequency 'passband'). In oversampled systems, with wideband digitizers, for very high data-rates, delays of tens or even hundreds of samples may go unnoticed, particularly if there are long lags in other system components, thus the filter delay provides an additional degree of design freedom that may be used to ensure system performance requirements are met.

The exponential weight is a limiting case of an Erlang weight. The un-normalized Erlang weight has the form of a uniformly sampled Erlang-distribution, which is simply an exponential

$e^{-m/\lambda_w} \equiv p^m$ multiplied by a 'monomial' (8a)

$m^\kappa$ i.e. (8b)

$w[m] = m^\kappa p^m$ where (8c)

$\kappa$ is an integer 'shape' parameter that is greater than or equal to zero and (8d)

$p = e^{-1/\lambda_w}$ (the 'smoothing' parameter) thus (8e)

$\lambda_w = -1/\ln p$ (the 'timescale' parameter, in samples). (8f)

Thus an exponential weight is a an Erlang weight with $\kappa = 0$ (see Figure 4).

The mean (or centroid), variance, and skew, of the Erlang weight (in continuous time) are

$\mu_w = (\kappa + 1)\lambda_w$ (8g)

$\sigma_w^2 = (\kappa + 1)\lambda_w^2$ and (8h)

$\eta_w = 2/\sqrt{\kappa + 1}$, respectively. (8i)

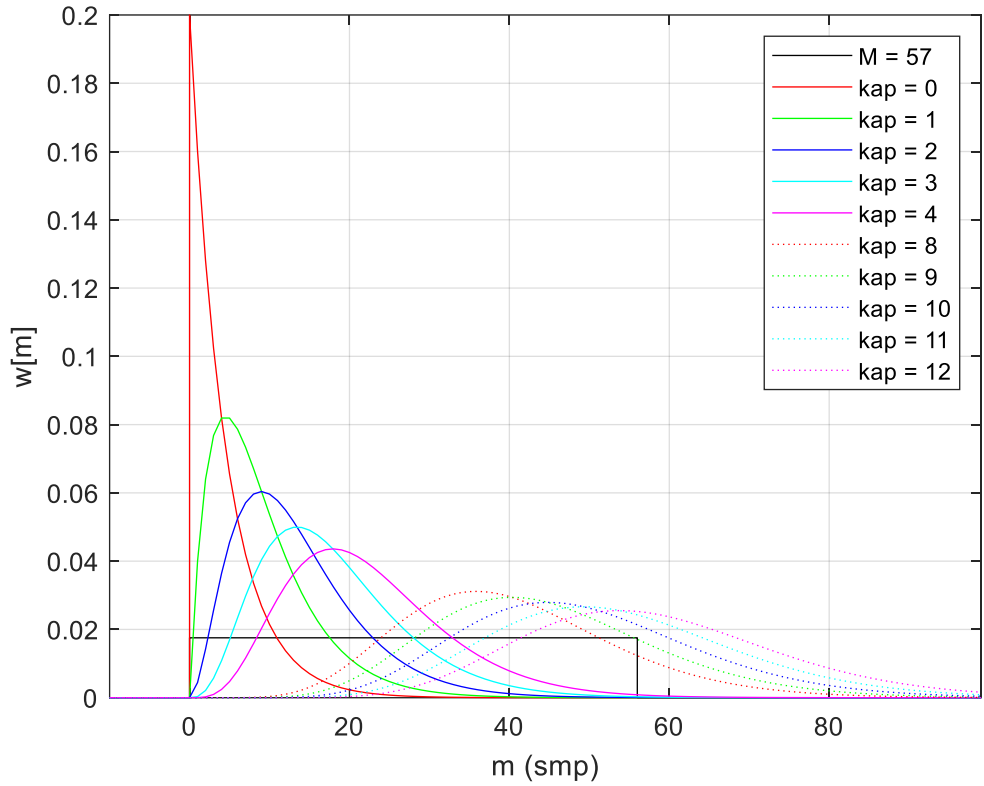

*Figure 4. Erlang weights (all with $p = 0.8$) for various shape parameters (i.e. κ, coloured lines) compared with a rectangular M-sample weight (black line). All windows are normalized for a unity sum. The length of the rectangular weight was set using (8h) so it has approximately the same variance as the Erlang weight with $\kappa = 12$.*

The digital circuit in Figure 5 consists of multiple leaky integrators in series and it may be used to recursively generate the impulse responses of the more general Erlang weights, i.e.

$w_k[m]$ for $k = 0 \dots \kappa$ where  (9a)

$w_k[m] = m^k p^m$  (9b)

or apply Erlang weights to a sample sequence $x[m]$ to compute moving averages. Such circuits are sometimes referred to as Laguerre networks because they may be used to recursively generate the discrete (associated) Laguerre functions or to convolve an input with such functions [15][16]. Unfortunately, the circuit as shown does not generate the desired moving (Erlang-weighted) averages directly however the desired quantities are simply obtained via a linear combination of its outputs. The coefficients required to combine its outputs will now be derived.

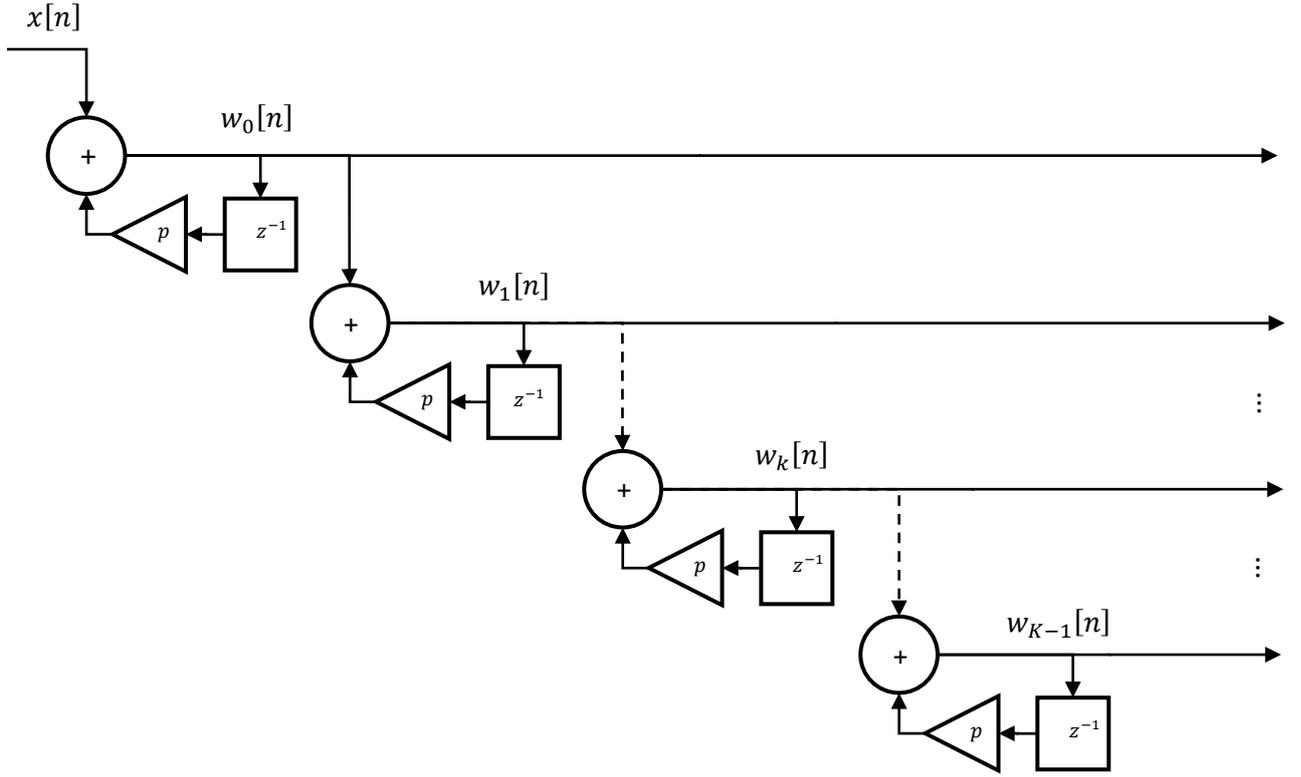

*Figure 5. A digital circuit of leaky integrators in series (i.e. a Laguerre network). The Erlang weights (i.e. $\mathscr{w}_k$) are formed from a linear combination of the internal states of this network (i.e. $w_k$).*

Linear (discrete-time) state-space notation provides an alternative, more convenient and compact, representation of this leaky integrating system, that is more amenable to analysis, extension and software implementation [4][11], i.e.

$\boldsymbol{w}[n] = \boldsymbol{G}\boldsymbol{w}[n-1] + \boldsymbol{H}x[n]$ where (10a)

$\boldsymbol{G}$ is a $K \times K$ lower-triangle matrix with $p$ along and below the diagonal and zeros above (10b)

$\boldsymbol{H}$ is a $K \times 1$ vector of ones and (10c)

$\boldsymbol{w}[n]$ is a $K \times 1$ vector that holds the $K$ internal states of the $K$th-order averaging system at the time of the $n$th sample, i.e. $w_k[n]$ for $k = 0 \dots K-1$, with $K = \kappa + 1$; (10d)

e.g. for $\kappa = 3$ such that $K = 4$ (i.e. a linear state-space system of fourth-order)

$$\boldsymbol{G} = \begin{bmatrix} p & 0 & 0 & 0 \\ p & p & 0 & 0 \\ p & p & p & 0 \\ p & p & p & p \end{bmatrix}, \boldsymbol{H} = \begin{bmatrix} 1 \\ 1 \\ 1 \\ 1 \end{bmatrix} \text{ and } \boldsymbol{w}[n] = \begin{bmatrix} w_0[n] \\ w_1[n] \\ w_2[n] \\ w_3[n] \end{bmatrix}.$$ (10e)

The impulse responses $\phi_k[m]$ that link the input to the states (i.e. $x[n]$ to $w_k[n]$) are a linear combination of $\mathscr{w}_k[m]$ thus the Erlang moving averages (for $k = 0 \dots \kappa$) of the input sequence $x[n]$ are formed from a linear combination of the internal states $w_k[n]$, i.e.

$\boldsymbol{y}[n] = \boldsymbol{C}\boldsymbol{w}[n]$ where (11a)

$C$ is a $K \times K$ lower triangle 'output' matrix (i.e. with zeros above the diagonal) and (11b)

$y[n]$ is a $K \times 1$ vector with the $k$th element $y_k[n]$ equal to the Erlang moving average computed using $w_k[m] = m^k p^m$ at the time of the $n$th sample. (11c)

Note that if only the $w[m] = m^\kappa p^m$ weighted average is required then only the ultimate row of $C$ is used (i.e. $k = K - 1$).

The elements of the output matrix are determined using

$C = \gamma T_{w \leftarrow \phi}$ where (12a)

$\gamma$ is a $K \times K$ matrix of zeros with normalizing factors $\gamma_k$ along the diagonal with (12b)

$\gamma_k = 1/\sum_{m=0}^{\infty} w_k[m]$. (12c)

$T_{w \leftarrow \phi}$ is a $K \times K$ lower triangle matrix (i.e. with zeros above the diagonal) that transforms the state impulse response into the Erlang weights, with elements $T_{w \leftarrow \phi}$ such that (12d)

$w_{k_a}[m] = \sum_{k_b=0}^{k_a} T_{w \leftarrow \phi}[k_a, k_b] \phi_{k_b}[m]$ for $k_a = 0 \ldots K - 1$. (12e)

The diagonal elements of $\gamma$ (i.e. normalizing factors $\gamma_k$) may be computed from the infinite summations

$S_k(p) = \sum_{m=0}^{\infty} p^m m^k$ provided in Table 1 (for $k = 0 \ldots 10$) using (13a)

$\gamma_k = 1/S_k(p)$. (13b)

Note that notation and analysis are simplified somewhat if normalizing factors are applied in the final stages of processing, rather than the initial stages, i.e. in $H$. The elements of the transformation matrix $T_{w \leftarrow \phi}$ are provided in

Table 2 for $k = 0 \ldots 5$. The impulse responses $\phi_k[m]$ (upper row), generated using the recursion above for a unit impulse input, and the un-normalized weights $w_k[m]$ (lower row) produced after $\boldsymbol{T}_{w \leftarrow \phi}$ has been used to transform the states, are plotted in Figure 6 for $k = 0 \ldots 3$ (left to right). Note also that for the exponential moving average, i.e. for $\kappa = 0$ thus $K = 1$, we have $\boldsymbol{G} = p$, $\boldsymbol{H} = 1$ & $\boldsymbol{C} = 1 - p$.

Table 1. Infinite summations of exponentially windowed monomials $p^m m^k$ or Erlang weights $p^m m^\kappa$ as a function of the smoothing parameter p.

| k | $S_k(p)$ |
|---|---|
| 0 | $1/(1-p)$ |
| 1 | $p/(1-p)^2$ |
| 2 | $(p^2 + p)/(1-p)^3$ |
| 3 | $(p^3 + 4p^2 + p)/(1-p)^4$ |
| 4 | $(p^4 + 11p^3 + 11p^2 + p)/(1-p)^5$ |
| 5 | $(p^5 + 26p^4 + 66p^3 + 26p^2 + p)/(1-p)^6$ |
| 6 | $(p^6 + 57p^5 + 302p^4 + 302p^3 + 57p^2 + p)/(1-p)^7$ |
| 7 | $(p^7 + 120p^6 + 1191p^5 + 2416p^4 + 1191p^3 + 120p^2 + p)/(1-p)^8$ |
| 8 | $(p^8 + 247p^7 + 4293p^6 + 15619p^5 + 15619p^4 + 4293p^3 + 247p^2 + p)/(1-p)^9$ |
| 9 | $(p^9 + 502p^8 + 14608p^7 + 88234p^6 + 156190p^5 + 88234p^4 + 14608p^3 + 502p^2 + p)/(1-p)^{10}$ |
| 10 | $(p^{10} + 1013p^9 + 47840p^8 + 455192p^7 + 1310354p^6 + 1310354p^5 + 455192p^4 + 47840p^3 + 1013p^2 + p)/(1-p)^{11}$ |

Table 2. Elements of the transform matrix (i.e. $T_{w\leftarrow\phi}$) that are used to populate the output matrix (i.e. $C$) which combines the internal states of the Laguerre network to form the Erlang weights.

| 1  | 0   | 0    | 0   | 0    | 0   |
|----|-----|------|-----|------|-----|
| -1 | 1   | 0    | 0   | 0    | 0   |
| 1  | -3  | 2    | 0   | 0    | 0   |
| -1 | 7   | -12  | 6   | 0    | 0   |
| 1  | -15 | 50   | -60 | 24   | 0   |
| -1 | 31  | -180 | 390 | -360 | 120 |

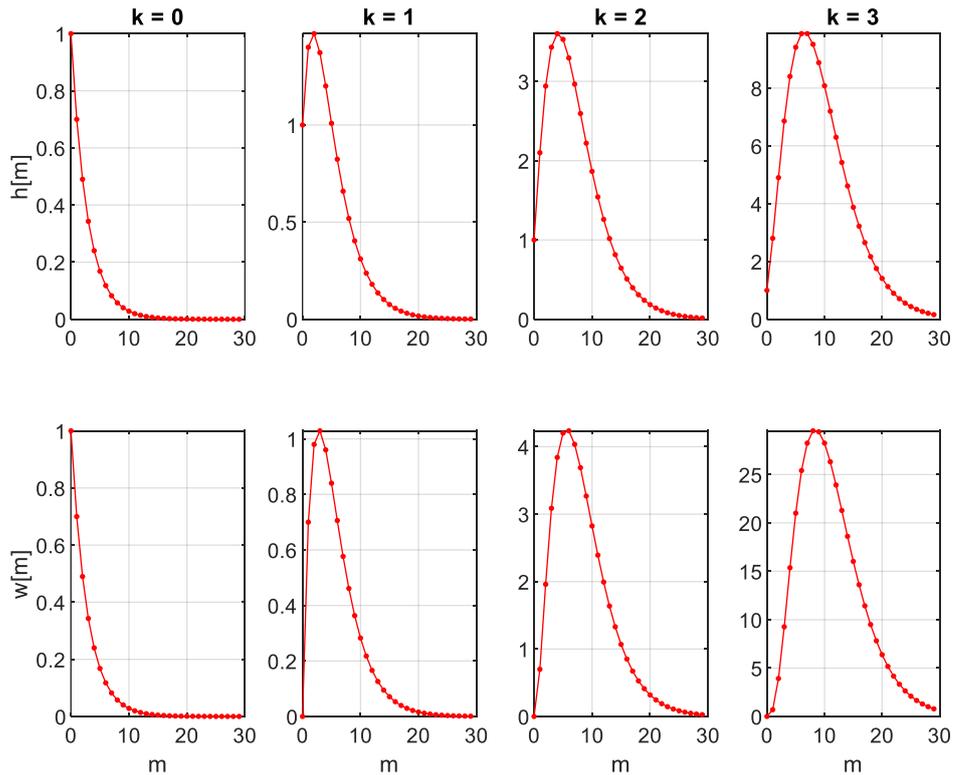

Figure 6. Impulse responses of the Laguerre network (i.e. $\phi_k[m]$, top row) and the (un-normalized) Erlang weights (i.e. $w_k[m]$, bottom row)

To reduce the impact of the filter start-up transient when computing recursive moving averages, the linear state-space recursion above should be initialized using

$w[0] = \rho x[0]$ where (14a)

$\rho$ is a constant vector with elements that are a function of $p$. (14b)

The elements of $\rho$ are set equal to the state vector $w[n]$ at the steady-state limit, i.e. as $n \to \infty$, for a unit step input with

$$x[n] = \begin{cases} 0, & n < 0 \\ 1, & n \geq 0 \end{cases}.$$  (15a)

The limit is readily found analytically by applying the final-value theorem to the discrete-time transfer function

$\mathcal{H}_{w_k \leftarrow x}(z) = W_k(z)/\mathcal{X}(z)$ that links $x[n]$ to $w_k[n]$. (16a)

Following the working above in (7), for $K = 1$ (and $k = 0$):

$$w_0[n] = pw_0[n-1] + x[n] \text{ thus} \tag{17a}$$

$$\mathcal{H}_{w_0 \leftarrow x}(z) = W_0(z)/\mathcal{X}(z) = z/(z-p) \text{ and} \tag{17b}$$

$$W_0(z) = z/(z-p)\,\mathcal{X}(z), \text{ or for } k \text{ of the leaky integrator elements in series:} \tag{17c}$$

$$W_k(z) = \{z/(z-p)\}^k\,\mathcal{X}(z). \tag{17d}$$

Now, for a unit step input  (18a)

$$\mathcal{X}(z) = z/(z-1), \text{ thus} \tag{18b}$$

$$W_k(z) = \{z/(z-p)\}^k\{z/(z-1)\}. \tag{18c}$$

The final value theorem [4], asserts that

$$\lim_{n \to \infty} w_k[n] = (z-1)W_k(z)|_{z=1} \text{ thus} \tag{19a}$$

$$\rho_k = 1/(1-p)^k \text{ where} \tag{19b}$$

$\rho_k$ is the $k$th element of $\boldsymbol{\rho}$, for $k = 0 \ldots K-1$. (19c)

Alternatively, the elements of $\boldsymbol{\rho}$ may be determined numerically by applying the recursion for a unit step input (i.e. $x[n] = 1$ for $n \geq 0$) until the steady-state regime is reached and the internal states of the leaky integrating network converge on their final values, to within a specified tolerance.

In summary, the use of Erlang weights in moving averages confers response characteristics that may be desirable in some applications, for instance: Relative to the simple moving average with a rectangular weight, a gradual taper yields a smoother response, with no sudden jumps in time and no nulls (or sidelobes) in frequency; furthermore, its recursive realization, using a bank of leaky integrators, provides rounding-error robustness on finite-precision machines. Relative to the exponential moving average, the two-sided taper of Erlang weights (with a non-zero shape parameter) increases the symmetry of the response in time and improves passband phase-linearity in frequency.

However, computing the mean (i.e. the magnitude of a direct-current component) of a sample sequence is a very rudimentary form of signal/data analysis. Fortunately, with only a few minor modifications/extensions, the basic filter-bank presented above may also be used for more sophisticated forms of online statistical analysis for non-stationary signals in non-Gaussian (white) noise. The working below shows that the state 'update' equations (i.e. $\boldsymbol{G}$ & $\boldsymbol{H}$) are unchanged and that only the state 'output' equations (i.e. $\boldsymbol{C}$) require modification. The derivation begins with a brief overview of linear least-squares regression with an arbitrary error weight, for moving/sliding statistical analysis in online systems. A finite $M$-sample weight and a non-recursive formulation is used initially to introduce the basic concepts in a familiar least-squares context. Exponential and Erlang weights of infinite extent (for a fading memory and a lagged fading memory, respectively) are then considered in recursive realizations.

# Section 2. Recursive weighted least-squares regression with polynomial models

Digital circuits with one or more (pure or leaky) integrating elements are used above to estimate the mean of an input sequence, i.e. the value of a single unknown parameter $X$, using multiple observations of a random variable $Y$, with

$Y[n] = X + \varepsilon[n]$ where (20a)

$\varepsilon$ is an uncorrelated random error,
with an expected value of zero and a variance of $\sigma_\varepsilon^2$, i.e. (20b)

$E\langle Y \rangle = X$ (20c)

$E\langle \varepsilon \rangle = 0$ (20d)

$\text{Cov}\langle \varepsilon[n_a], \varepsilon[n_b] \rangle = 0$ for $n_a \neq n_b$ and (20e)

$\text{Cov}\langle \varepsilon[n_a], \varepsilon[n_b] \rangle = \text{Var}\langle \varepsilon \rangle = \sigma_\varepsilon^2$ for $n_a = n_b$ where (20f)

$E\langle \blacksquare \rangle$ is the expectation operator (20g)

(note that the random error is 'white' but it is not necessarily Gaussian) [10].

For this first-order model with a single degree of freedom (i.e. $K_X = 1$) the expected value of the random variable is equal to the mean, which is a constant value and independent of time thus the sample index $n$, i.e.

$E\langle Y \rangle = X = \alpha_0$. (21a)

For more general $K_X$th-order linear models with $K_X$ degrees of freedom, applied via convolution operations (i.e. with digital filters) $X$ is a function of the delay index $m$ and we have

$Y[n-m] = X[m] + \varepsilon[n]$ where (22a)

$X[m] = \sum_{k_X=0}^{K_X-1} \alpha_{k_X} \varphi_{k_X}[m]$ with (22b)

$\varphi_{k_X}[m] = m^{k_X}$ for a polynomial model with $K_X$ monomial terms and
a polynomial with a degree of $K_X - 1$, thus (22c)

$E\langle Y[n-m] \rangle = X[m] = \alpha_0 + \alpha_1 m \ldots + \alpha_{k_X} m^{k_X} \ldots + \alpha_{K-1} m^{K_X-1}$. (22d)

Note that for sliding (i.e. moving) regression analysis, it is convenient to place the origin of the model at the time of the most recent sample, i.e. at $n$ (for $n = 0 \ldots \infty$) where $m = 0$.

For a given model and measurement sequence, the weighted sum-of-squared residuals is defined as

$S_\varepsilon = \sum_{m=0}^{M-1} w[m](Y[n-m] - X[m])^2$. (23a)

It will initially be assumed that $w[m]$ is of finite length, i.e. $M < \infty$, e.g. a uniform weight, and it is not assumed to be normalized. The weight need not be constant over the windowed interval, for instance a Gaussian weight is recommended for the multidimensional filters in [17]. Finite windows lead to the standard weighted least-squares formulation; however, for large $M$ this leads to FIR realizations of high computational complexity. We begin with finite weights below because they simplify the theory and lead to recursive IIR realizations, with reduced computational complexity, when Erlang weights with $M = \infty$ are used.

The unknown parameters of a model may be estimated using the principle of least squares [10][13], i.e. by determining the parameter values $\alpha_{k_X}$ that minimize $S_\varepsilon$. This is done by differentiating $S_\varepsilon$ with respect to $\alpha_{k_X}$, setting the $K_X$ partial derivatives to zero, then solving for $\alpha_{k_X}$. Using matrix notation and a $K_X$th-order polynomial model, with $K_X$ monomial components, thus a polynomial degree of $K_X - 1$, we have the standard result:

$$\boldsymbol{\alpha} = \left(\boldsymbol{X}_\varphi^T \boldsymbol{\mathcal{W}} \boldsymbol{X}_\varphi\right)^{-1} \boldsymbol{X}_\varphi^T \boldsymbol{\mathcal{W}} \boldsymbol{y} \text{ and} \tag{24a}$$

$$\boldsymbol{x} = \boldsymbol{X}_\varphi \boldsymbol{\alpha} \text{ where} \tag{24b}$$

$\boldsymbol{\alpha}$ is a $K_X \times 1$ 'coefficient' vector with elements $\hat{\alpha}_{k_X}$, for $k_X = 0 \ldots K_X - 1$,
i.e. the estimated model parameters (24c)

$\boldsymbol{y}$ is an $M \times 1$ 'data' vector with the $m$th element equal to $Y[n-m]$,
for $m = 0 \ldots M - 1$, i.e. the $M$ most recent observations (newest to oldest) (24d)

$\boldsymbol{X}_\varphi$ is an $M \times K_X$ 'regressor' matrix
with elements $\varphi_{k_X}[m] = m^{k_X}$ in the $m$th row and $k_X$th column (24e)

$\boldsymbol{\mathcal{W}}$ is an $M \times M$ 'weight' matrix of zeros with $\mathcal{W}[m,m] = w[m]$ along the diagonal (24f)

$\boldsymbol{x}$ is an $M \times 1$ vector with the $m$th element equal to $X[m]$ (24g)

$\blacksquare^T$ is the transpose operator (24h)

$\blacksquare^{-1}$ is the matrix inverse operator. (24i)

Or equivalently:

$$\boldsymbol{\alpha} = \boldsymbol{S}_{\varphi\mathcal{W}\varphi}^{-1} \boldsymbol{s}_{\varphi\mathcal{W}Y} \text{ where} \tag{25a}$$

$$\boldsymbol{S}_{\varphi\mathcal{W}\varphi} = \boldsymbol{X}_\varphi^T \boldsymbol{\mathcal{W}} \boldsymbol{X}_\varphi \text{ (a } K_X \times K_X \text{ weighted 'overlap' matrix) with elements} \tag{25b}$$

$$S_{\varphi\mathcal{W}\varphi}[k_a, k_b] = \sum_{m=0}^{M-1} \varphi_{k_a}[m] w[m] \varphi_{k_b}[m] \text{ in the } k_a\text{th row and } k_b\text{th column} \tag{25c}$$

$$\boldsymbol{s}_{\varphi\mathcal{W}Y} = \boldsymbol{X}_\varphi^T \boldsymbol{\mathcal{W}} \boldsymbol{y} \text{ (a } K_X \times 1 \text{ weighted 'projection' vector) with elements} \tag{25d}$$

$$s_{\varphi\mathcal{W}Y}[k] = \sum_{m=0}^{M-1} \varphi_{k_X}[m] w[m] Y[n-m], \text{ for } k_X = 0 \ldots K_X - 1. \tag{25e}$$

The weighted sum-of-squared residuals for the fitted model is evaluated by substituting $\boldsymbol{x} = \boldsymbol{X}_\varphi \boldsymbol{\alpha}$ and $\boldsymbol{y}$ in (24g) and (24d) above, for $X$ and $Y$ in (23a) which yields

$$S_\varepsilon = \boldsymbol{y}^T \boldsymbol{\mathcal{W}} \boldsymbol{y} - 2\boldsymbol{\alpha}^T \boldsymbol{X}_\varphi^T \boldsymbol{\mathcal{W}} \boldsymbol{y} + \boldsymbol{\alpha}^T \boldsymbol{X}_\varphi^T \boldsymbol{\mathcal{W}} \boldsymbol{X}_\varphi \boldsymbol{\alpha} \tag{26a}$$

or equivalently, using the definitions of $\boldsymbol{S}_{\varphi\mathcal{W}\varphi}$ and $\boldsymbol{s}_{\varphi\mathcal{W}Y}$ given in (25b) & (25d):

$$S_\varepsilon = s_{Y\mathcal{W}Y} - 2\boldsymbol{\alpha}^T \boldsymbol{s}_{\varphi\mathcal{W}Y} + \boldsymbol{\alpha}^T \boldsymbol{S}_{\varphi\mathcal{W}\varphi} \boldsymbol{\alpha} \text{ where} \tag{26b}$$

$$s_{Y\mathcal{W}Y} = \boldsymbol{y}^T \boldsymbol{\mathcal{W}} \boldsymbol{y}. \tag{26c}$$

The value of $S_\varepsilon$ provides an indication of the 'goodness' of fit and it may be used to estimate the variance of the measurement noise, using

$\hat{\sigma}_\varepsilon^2 = \gamma S_\varepsilon$ where (27a)

$\gamma$ is the normalizing factor for the applied weight, e.g. (27b)

$\gamma = 1/M$ for a rectangular weight. (27c)

When the model is inadequate or the system is disturbed, bias errors and random errors both contribute to $\hat{\sigma}_\varepsilon^2$. However, in most applications the distinction is moot, and it does not matter whether the error is white or coloured. Indeed, errors are almost always correlated (i.e. coloured) to some degree for real signals and systems.

Instead of using the *raw* polynomial model with monomial components $\varphi_{k_X}[m]$, it is easier to work with an *orthogonal* polynomial model with orthonormal components $\psi_{k_X}[m]$, that are normalized and orthogonal with respect to the nominated weight $w[m]$, so that the weighted overlap matrix is equal to the $K_X \times K_X$ identity matrix [1][2], i.e.

$\boldsymbol{S}_{\psi \mathcal{W} \psi} = \boldsymbol{X}_\psi^T \boldsymbol{\mathcal{W}} \boldsymbol{X}_\psi = \boldsymbol{I}_{K_X \times K_X}$ thus (28a)

$\boldsymbol{\beta} = \boldsymbol{X}_\psi^T \boldsymbol{\mathcal{W}} \boldsymbol{Y} = \boldsymbol{s}_{\psi \mathcal{W} Y}$ and (28b)

$S_\varepsilon = s_{Y \mathcal{W} Y} - 2\boldsymbol{\beta}^T \boldsymbol{s}_{\psi \mathcal{W} Y} + \boldsymbol{\beta}^T \boldsymbol{S}_{\psi \mathcal{W} \psi} \boldsymbol{\beta} = s_{Y \mathcal{W} Y} - \boldsymbol{\beta}^T \boldsymbol{\beta}$ where (28c)

$\boldsymbol{\beta}$ are the model coefficients such that (28d)

$X[m] = \sum_{k_X=0}^{K_X-1} \beta_{k_X} \psi_{k_X}[m]$ . (28e)

The $K_X$ orthonormal components $\psi_{k_X}[m]$ are a linear combination of the $K_X$ monomial components

$\varphi_{k_X}[m] = m^{k_X}$, i.e. (29a)

$\psi_{k_a}[m] = \sum_{k_b=0}^{k_a} T_{\psi \leftarrow \varphi}[k_a, k_b] \varphi_{k_b}[m]$ and (29b)

$\beta[k_a] = \sum_{k_b=0}^{k_a} T_{\psi \leftarrow \varphi}[k_a, k_b] \alpha[k_b]$ for $k_a = 0 \ldots K_X - 1$ where (29c)

$T_{\psi \leftarrow \varphi}[k_a, k_b]$ are elements of the ($K_X \times K_X$ lower-triangle)
'transformation' matrix $\boldsymbol{T}_{\psi \leftarrow \varphi}$, with (29d)

$\boldsymbol{X}_\psi = \boldsymbol{X}_\varphi \boldsymbol{T}_{\varphi \leftarrow \psi}$ where (29e)

$\boldsymbol{T}_{\varphi \leftarrow \psi} = \boldsymbol{T}_{\psi \leftarrow \varphi}^T$ . (29f)

Thus, we seek a ($K_X \times K_X$ upper-triangle) transformation matrix $\boldsymbol{T}_{\varphi \leftarrow \psi}$, such that

$\boldsymbol{S}_{\psi \mathcal{W} \psi} = \boldsymbol{X}_\psi^T \boldsymbol{\mathcal{W}} \boldsymbol{X}_\psi = \boldsymbol{T}_{\varphi \leftarrow \psi}^T \boldsymbol{X}_\varphi^T \boldsymbol{\mathcal{W}} \boldsymbol{X}_\varphi \boldsymbol{T}_{\varphi \leftarrow \psi} = \boldsymbol{T}_{\varphi \leftarrow \psi}^T \boldsymbol{S}_{\varphi \mathcal{W} \varphi} \boldsymbol{T}_{\varphi \leftarrow \psi} = \boldsymbol{I}_{K_X \times K_X}$ . (30a)

The required transform is found via a Cholesky factorization of $\boldsymbol{S}_{\varphi \mathcal{W} \varphi}$ such that

$\boldsymbol{S}_{\varphi \mathcal{W} \varphi} = \boldsymbol{L} \boldsymbol{U}$ where (31a)

$\boldsymbol{L}$ and $\boldsymbol{U}$ are lower and upper triangle matrices with $\boldsymbol{L} = \boldsymbol{U}^T$. Thus (31b)

$\boldsymbol{L}^{-1} \boldsymbol{S}_{\varphi \mathcal{W} \varphi} \boldsymbol{U}^{-1} = \boldsymbol{I}_{K_X \times K_X}$ and the transformation matrix is (32a)

$\boldsymbol{T}_{\varphi \leftarrow \psi} = \boldsymbol{U}^{-1}$ with (32b)

$\boldsymbol{T}_{\varphi \leftarrow \psi}^T = \boldsymbol{L}^{-1} = \boldsymbol{T}_{\psi \leftarrow \varphi}$; note also that (32c)

$$S_{\varphi W \varphi}^{-1} = T_{\varphi \leftarrow \psi} T_{\psi \leftarrow \varphi} \tag{32d}$$

$$X_\varphi = X_\psi T_{\psi \leftarrow \varphi} \tag{32e}$$

$$\varphi_{k_a}[m] = \sum_{k_b=k_a}^{K_X-1} T_{\varphi \leftarrow \psi}[k_a, k_b] \psi_{k_a}[m] \text{ and} \tag{32f}$$

$$\alpha_{k_a} = \sum_{k_b=k_a}^{K_X-1} T_{\varphi \leftarrow \psi}[k_a, k_b] \beta_{k_b} \text{ for } k_a = 0 \ldots K_X - 1. \tag{32g}$$

For rectangular weights with a *finite memory* (28b) & (28c) may be evaluated as a batch or recursively using a bank of lossless integrators to compute the required running sums. In online systems, the former approach inefficient whereas the latter approach is efficient but (potentially) imprecise; thus, the finite-memory case is not considered further here.

For Erlang weights with a *fading memory* or a *lagged fading memory* (for $\kappa = 0$ and $\kappa > 0$, respectively) infinite summations replace the finite summations in (25c) and (25e) to evaluate the elements of $S_{\varphi W \varphi}$ and $s_{\varphi WY}$.

Instead of (weighted) vector dot products, the (constant) elements of $S_{\varphi W \varphi}$ are now set using the infinite sums

$$S_{\varphi W \varphi}[k_a, k_b] = S_{k_a + k_b + \kappa}(p) \text{ where} \tag{33a}$$

$S_k(p)$ is defined above in (13a) and evaluated in Table 1. \hfill (33b)

For a given value of $p$, approximate values may also be found numerically on a computer by performing the summation using a loop until the sum converges to within a specified tolerance.

The (non-constant) elements of $s_{\varphi WY}$ at the time of the $n$th sample are efficiently (and precisely) computed using a bank of $K$ leaky integrators, via the linear state-space recursion in (10). Thus, the $k$th element of $s_{\varphi WY}$ at the time of the $n$th sample is set using $s_{\varphi WY}[k] = y_k[n]$ in (11) then, as indicated in (28b), (29c) & (32g) the coefficient vectors for the orthonormal and raw (i.e. non-orthogonal) polynomial models are simply found using

$$\boldsymbol{\beta} = T_{\psi \leftarrow \varphi} s_{\varphi WY} = s_{\psi WY} \text{ then} \tag{34a}$$

$$\boldsymbol{\alpha} = T_{\varphi \leftarrow \psi} \boldsymbol{\beta}. \tag{34b}$$

As indicated in (27a) & (28c), *two* parallel circuits or recursions are required to evaluate $S_\varepsilon$ for an estimate of the measurement noise variance using

$$\hat{\sigma}_\varepsilon^2 = \gamma S_\varepsilon = \gamma (s_{YWY} - \boldsymbol{\beta}^T \boldsymbol{\beta}) \text{ where} \tag{35a}$$

$\gamma$ is the normalizing factor for the weighting function. \hfill (35b)

For an Erlang weight

$$\gamma = 1/S_k(p) \text{ with} \tag{36a}$$

$$k = \kappa, \text{ where} \tag{36b}$$

$S_k(p)$ is defined above in (13a). \hfill (36c)

The *first* recursion computes the coefficient vector $\boldsymbol{\beta}$ with

$$E\langle Y[n-m]\rangle = \hat{X}[m] = X_\psi \boldsymbol{\beta}. \tag{37a}$$

It is a leaky integrating system of order $K_1 = \kappa + K_X$, i.e.

$\boldsymbol{w}_1[n] = \boldsymbol{G}_1\boldsymbol{w}_1[n-1] + \boldsymbol{H}_1 x_1[n]$ where (38a)

$\boldsymbol{G}_1$ and $\boldsymbol{H}_1$ are defined above in (10) using $K = K_1$ with (38b)

$\boldsymbol{y}_1[n] = \boldsymbol{C}_1\boldsymbol{w}_1[n]$ where (38c)

$\boldsymbol{y}_1[n] = \boldsymbol{\beta}$ (a $K_X \times 1$ vector). (38d)

The $K_X \times K_1$ output matrix for this recursion (i.e. $\boldsymbol{C}_1$) is defined as follows:

$\boldsymbol{C}_1 = \boldsymbol{T}_{\psi \leftarrow \varphi} \boldsymbol{T}_{\varphi \leftarrow w} \boldsymbol{T}_{w \leftarrow \phi}$ where (39a)

$\boldsymbol{T}_{w \leftarrow \phi}$ is the $K_1 \times K_1$ matrix defined above in (12) (39b)

$\boldsymbol{T}_{\varphi \leftarrow w} = [\boldsymbol{0}_{K_X \times \kappa}, \boldsymbol{I}_{K_X \times K_X}]$ (with $\boldsymbol{0}_{K_X \times \kappa}$ being a $K_X \times \kappa$ matrix of zeros) and (39c)

$\boldsymbol{T}_{\psi \leftarrow \varphi}$ is the $K_X \times K_X$ matrix defined above in (29). (39d)

For this linear state-space system the ■$_1$ subscript denotes the filtering of the sampled data sequence, raised to the power of one, i.e. $x_1[n] = x[n]$. This 'first-moment' recursion is initialized using

$\boldsymbol{w}_1[0] = \boldsymbol{\rho} x_1[n]$ where (40a)

$\boldsymbol{\rho}$ is the $K_1 \times 1$ steady-state vector defined above in (19). (40b)

The *second* recursion computes the expected value of the sampled data-sequence squared i.e.

$s_{YWY} = E\langle Y[n]^2 \rangle$. (41a)

It is a leaky integrating system of order $K_2 = \kappa + 1$, i.e.

$\boldsymbol{w}_2[n] = \boldsymbol{G}_2\boldsymbol{w}_2[n-1] + \boldsymbol{H}_2 x_2[n]$ where (41b)

$\boldsymbol{G}_2$ and $\boldsymbol{H}_2$ are defined above in (10) using $K = K_2$ with (41c)

$y_2[n] = \boldsymbol{C}_2\boldsymbol{w}_2[n]$ where (41d)

$y_2[n] = s_{YWY}$ (a $1 \times 1$ scalar) and (41e)

$\boldsymbol{C}_2$ is a row vector of length $K_2$ equal to the ultimate row of $\boldsymbol{T}_{w \leftarrow \phi}$ with elements
$C_2[k_2] = \boldsymbol{T}_{w \leftarrow \phi}[K_2 - 1, k_2]$ for $k_2 = 0 \ldots K_2 - 1$. (41f)

For this linear state-space system the ■$_2$ subscript denotes the filtering of the sampled data sequence, raised to the power of two, i.e. $x_2[n] = x[n]x[n]$. This 'second-moment' recursion is initialized using

$\boldsymbol{w}_2[0] = \boldsymbol{\rho}(\sigma_0^2 + x_2[0])$ where (42a)

$\boldsymbol{\rho}$ is the $K_2 \times 1$ steady-state vector defined above in (19) and (42b)

$\sigma_0^2$ is a coarse estimate of the measurement error variance, i.e. $\sigma_\varepsilon^2$. (42c)

Note that for an Erlang weight and monomial components,

$w_\kappa[m]\varphi_{k_X}[m] = w_0[m]\varphi_{k_X+\kappa}[m] = w_{k_X+\kappa}[m] = m^{k_X+\kappa} p^m$ (43a)

thus $G$ and $H$ for the leaky integrating system may be used for all combinations of $\kappa$ & $K_X$ and it does not matter whether the powers of $m$ 'belong' to the error weight or the model component. A distinction is only made when the $C$ operator is formed.

## Section 3. Sequential smoothing and the estimation of derivative states

For a rectangular weight (thus a finite memory) the $K_X$ orthonormal components $\psi_{k_X}[m]$ are the discrete *Legendre* polynomials [1][2]. For a simple exponential weight $w[m] = p^m$ (thus a fading memory) the discrete *Laguerre* polynomials result [1][2]. The more general Erlang weight $w[m] = m^\kappa p^m$ (with a lagged fading memory) delays the centroid of the weight and decreases its asymmetry (i.e. reduces its skew) for a recursive estimator with an increased phase lag but improved passband magnitude flatness and phase linearity [3]. This weight yields the discrete *associated Laguerre* polynomials. Use of the Legendre and Laguerre polynomials in sequential smoothing is detailed in Morrison's 1969 treatise; however, the associated Laguerre polynomials are not considered there.

Orthogonal polynomials simplify the derivation and realization of recursive regression for the various error weighting functions; however, the determination of the model coefficient vector $\boldsymbol{\beta}$, i.e. the orthogonal polynomial 'spectrum', or the monomial coefficient vector $\boldsymbol{\alpha}$, is rarely the objective of online data-analysis operations. Rather, information is usually inferred from the fitted model, such as the expected value of the observed random variable $Y$, at some point in the past, present or future (i.e. interpolation, smoothing and prediction) or its rate of change with respect to time (i.e. temporal derivatives) along with the corresponding covariance matrix $\boldsymbol{\Sigma}$, from which approximate error bounds may be derived.

In a Kalman filter recursion, the state covariance matrix is *derived* from prior knowledge of (process and measurement) noise variance and an initial covariance matrix for the state. It is assumed that the actual measurement-noise variance matches the provided measurement-noise variance, which is rarely the case in many applications. In the recursive regression filters considered here, the state covariance matrix is *estimated* using the dispersion of the observations around the fitted model, i.e. the weighted sum of squared residuals $S_\varepsilon$. Thus, the computed covariance matrix exhibits adaptive behaviour that is not usually found in a standard Kalman filter (i.e. a Kalman filter that does not incorporate probabilistic data association or multiple measurement models, etc.).

Recursive discrete-time filters to compute derivatives of low-frequency signals may also be designed via 'flatness constraints' in the $\omega$-domain [3]. Furthermore, recursive discrete-time observers to estimate the derivative states of $K_X$th-order integrating systems may also be designed via 'pole placement' in the $z$-domain [11]. However, these design procedures do not utilize the concepts of a 'fit' or an error 'weight' (which must be a non-negative function, by definition) thus they do not necessarily yield covariance matrices, which are shown to be very useful in Section 5.

According to the polynomial model [1][2], the expected value of the observed random variable at past ($m > 0$) present ($m = 0$) and future ($m < 0$) samples is evaluated using

$$E\langle Y[n-m]\rangle = X[m] = \sum_{k_X=0}^{K_X-1} \alpha_{k_X} m^{k_X} \text{ where} \tag{44a}$$

$\alpha_{k_X}$ is the $k_X$th element of $\boldsymbol{\alpha}$, determined using (34)

or for a sampling period of $T_s$ (seconds) and a 'relative time offset' of $\tau = -T_s m$ (seconds) in the past ($\tau < 0$) present ($\tau = 0$) or future ($\tau > 0$), relative to the current time of $t = T_s n$ (seconds)

$$E\langle Y(t+\tau)\rangle = X(\tau) = \sum_{k_X=0}^{K_X-1} \left(\frac{1}{-T_s}\right)^{k_X} \alpha_{k_X} \tau^{k_X}. \tag{45a}$$

The $k_t$th derivative of $X(\tau)$ with respect to time (for $k_t = 0 \ldots K_t - 1$, where $K_t \leq K_X$) is therefore

$$X^{(k_t)}(\tau) = \sum_{k_X=0}^{K_X-1} \left(\frac{1}{-T_s}\right)^{k_X} \frac{k_X!}{(k_X-k_t)!} \alpha_{k_X} \tau^{k_X-k_t} \text{ where} \tag{46a}$$

$$X^{(k_t)}(\tau) = \frac{d^{k_t}X(\tau)}{d\tau^{k_t}} \text{ thus} \tag{46b}$$

$$\hat{X}^{(k_t)}(\tau) = \sum_{k_X=0}^{K_X-1} \left(\frac{1}{-T_s}\right)^{k_X} \frac{k_X!}{(k_X-k_t)!} \hat{\alpha}_{k_X} \tau^{k_X-k_t} \text{ where} \tag{46c}$$

$\hat{X}^{(k_t)}(\tau)$ is the weighted least-squares estimate of $X^{(k_t)}(\tau)$. $\tag{46d}$

The desired estimates of the derivatives are therefore computed using the linear state-space equations in (38) with an alternative output operator using:

$$\boldsymbol{w}_1[n] = \boldsymbol{G}_1 \boldsymbol{w}_1[n-1] + \boldsymbol{H}_1 x_1[n] \text{ and} \tag{47a}$$

$$\boldsymbol{y}_q[n] = \boldsymbol{C}_q \boldsymbol{w}_1[n] \text{ where} \tag{47b}$$

$\boldsymbol{y}_q[n]$ is a vector of length $K_t$ with the $k_t$th element equal to $\hat{X}^{(k_t)}(\tau)$,
evaluated at a constant nominated time offset of $\tau = -qT_s$ and $\tag{47c}$

$\boldsymbol{C}_q$ is the $K_t \times K_1$ output matrix with $\boldsymbol{C}_q = \boldsymbol{\mathcal{D}}_q \boldsymbol{T}_{\varphi \leftarrow \psi} \boldsymbol{C}_1$. $\tag{47d}$

In the definitions above, $q$ is the 'delay' parameter (with a non-integer value, in samples) and $\boldsymbol{\mathcal{D}}_q$ is a $K_t \times K_X$ 'synthesis' matrix, that differentiates and evaluates the fitted polynomial model at time $t = (n-q)T_s$, with elements

$$\mathcal{D}_q[k_t, k_X] = \begin{cases} \left(\frac{1}{-T_s}\right)^{k_t} \frac{k_X!}{(k_X-k_t)!} q^{k_X-k_t}, & k_t \leq k_X \\ 0, & k_t > k_X \end{cases} \text{ for} \tag{48a}$$

$k_t = 0 \ldots K_t - 1$ and $k_X = 0 \ldots K_X - 1$. $\tag{48b}$

As indicated in (47d) the modified output operator ($\boldsymbol{C}_q$) computes the filter outputs in (47b) from the feedback section in (47a) by combining three operations into a single linear combination. Firstly, $\boldsymbol{C}_1$ transforms the ($K_1 \times 1$) vector of internal states ($\boldsymbol{w}_1$) into a ($K_X \times 1$) vector of coefficient estimates ($\hat{\beta}_{k_X}$) for the orthogonal polynomial model, see (39a). Secondly, $\boldsymbol{T}_{\varphi \leftarrow \psi}$ transforms this vector into a ($K_X \times 1$) vector of coefficient estimates ($\hat{\alpha}_{k_X}$) for the non-orthogonal monomial model, see (29)-(32); then finally, $\boldsymbol{\mathcal{D}}_q$ converts these coefficients into a vector of derivative estimates, i.e. $\hat{X}^{(k_t)}(\tau)$, by differentiating and evaluating the fitted polynomial at the specified point in time, as set using $q$, e.g. $q = 0.5$ corresponds to a time that is halfway between the current and previous samples, for retrospective interpolation; $q = -1.0$ corresponds to a time that is one sample ahead in the future, for prediction of the next input, etc.

Now let $\hat{\sigma}_{k_t}^2$ be the variance of the $k_t$th-order derivative estimate,

$\hat{\sigma}_{k_t}^2 = \text{Var}\langle \hat{X}^{(k_t)}(\tau) - X^{(k_t)}(\tau) \rangle$, i.e. the expected squared error at steady state; and let $\tag{49a}$

$\boldsymbol{\Sigma}$ be the ($K_t \times K_t$) covariance matrix of the estimate with elements $\tag{49b}$

$\Sigma[k_a, k_b] = \text{Cov}\langle \hat{X}^{(k_a)}(\tau) - X^{(k_a)}(\tau), \hat{X}^{(k_b)}(\tau) - X^{(k_b)}(\tau) \rangle$. $\tag{49c}$

Intuitively, it is reasonable to expect $\hat{\sigma}^2_{k_t}$ and the elements of $\mathbf{\Sigma}$ to:

a) increase with the estimate of the measurement-noise variance, i.e. $\hat{\sigma}^2_\varepsilon$;
b) increase with the order of the temporal derivatives, i.e. $k_t$ and
c) be higher at relative time offsets with reduced data 'support', i.e. far from the centroid of the weight, e.g. in the future or in the distant past.

Indeed, $\hat{\sigma}^2_{k_t}$ is proportional to $\hat{\sigma}^2_\varepsilon$ and the constant of proportionality is sometimes referred to as the variance reduction factor (VRF, in the time domain) or the white-noise gain (WNG, in the frequency domain). The VRF (a dimensionless factor) quantifies the attenuation (VRF $< 1$) or amplification (VRF $> 1$) of white noise power as it passes through the state estimator. Let $h_{k_t}[m]$ and $H_{k_t}(\omega)$ be the impulse response and the frequency response (respectively) of the estimator's discrete-time transfer function that links the input $x[n]$ to the $k_t$th element of the output vector $\mathbf{y}_q[n]$. In the discrete time domain, the output of the estimator is determined by convolving the white-noise input sequence $x[n]$, which has an average power of $\hat{\sigma}^2_\varepsilon$, with the impulse response $h_{k_t}[m]$. In the frequency domain, the white-noise input has a constant power spectral density of $\hat{\sigma}^2_\varepsilon$ thus this convolution is equal to the product of $\hat{\sigma}_\varepsilon$ and $H_{k_t}(\omega)$; therefore, the power of the output sequence is

$$\hat{\sigma}^2_{k_t} = \frac{\hat{\sigma}^2_\varepsilon}{2\pi} \int_{-\pi}^{\pi} |H_{k_t}(\omega)|^2 d\omega \ . \tag{50a}$$

According to Parseval's theorem, this definite integral may be evaluated in the discrete-time domain via the infinite summation

$$\hat{\sigma}^2_{k_t} = \hat{\sigma}^2_\varepsilon \sum_{m=0}^{\infty} |h_{k_t}[m]|^2 \text{ thus} \tag{51a}$$

$$\text{VRF} = \sum_{m=0}^{\infty} |h_{k_t}[m]|^2 \ . \tag{51b}$$

Similarly, $\mathbf{\Sigma}$ is set using the outer product of inner products

$$\mathbf{\Sigma}[k_a, k_b] = \hat{\sigma}^2_\varepsilon \sum_{m=0}^{\infty} |h_{k_a}[m] h_{k_b}[m]| \text{ or} \tag{52a}$$

$$\mathbf{\Sigma} = \hat{\sigma}^2_\varepsilon \mathbf{C}_\varphi \mathbf{S}_{\mathcal{W}\varphi\mathcal{W}\varphi} \mathbf{C}_\varphi^T = \hat{\sigma}^2_\varepsilon \mathbf{VRF} \text{ where} \tag{52b}$$

$$\mathbf{VRF} = \mathbf{C}_\varphi \mathbf{S}_{\mathcal{W}\varphi\mathcal{W}\varphi} \mathbf{C}_\varphi^T \text{ is a } (K_t \times K_t) \text{ matrix of (co-)variance reduction factors} \tag{52c}$$

$$\mathbf{S}_{\mathcal{W}\varphi\mathcal{W}\varphi} = \mathbf{W}^T \mathbf{X}_\varphi^T \mathbf{W} \mathbf{X}_\varphi \text{ is a } (K_X \times K_X) \text{ matrix of infinite summations with elements} \tag{52c}$$

$$S_{\mathcal{W}\varphi\mathcal{W}\varphi}[k_a, k_b] = \sum_{m=0}^{\infty} w[m] \varphi_{k_a}[m] w[m] \varphi_{k_b}[m]$$
which have the same form as the $S_{\varphi\mathcal{W}\varphi}[k_a, k_b]$ summations
used previously in (33); although, the weight is now included twice, i.e. $\tag{52d}$

$$S_{\mathcal{W}\varphi\mathcal{W}\varphi}[k_a, k_b] = \sum_{m=0}^{\infty} p^m m^{k_a+\kappa} p^m m^{k_b+\kappa} = \mathcal{S}_{k_a+k_b+2\kappa}(p^2) \text{ (see Table 1) and} \tag{52e}$$

$$\mathbf{C}_\varphi = \mathbf{D}_q \mathbf{T}_{\varphi \leftarrow \psi} \mathbf{T}_{\psi \leftarrow \varphi} = \mathbf{D}_q \mathbf{S}_{\varphi\mathcal{W}\varphi}^{-1} \ . \tag{52f}$$

The estimate of the measurement error variance $\hat{\sigma}^2_\varepsilon$ is a 'live' measure of data dispersion around the least-squares fitted polynomial and it is updated for each new sample. This factor scales the **VRF** matrix, which is fixed on design and is determined by the selected regression parameters, i.e. the error weight (set using $\kappa$ & $p$), the model order ($K_X$) and the time at which the least-squares fitted polynomial is evaluated (set using $q$). In principle, the polynomial may be evaluated at any point in time; however, if the model order is large and the support of the weight is insufficient, then large errors (as indicated by the elements of the **VRF** matrix) should be expected, even when the measurement-noise power is low (as indicated by $\hat{\sigma}_\varepsilon^2$). For example, the variance reduction factor increases rapidly for predictions (set using $q < 0$).

Clearly the **VRF** matrix is a function of the relative delay parameter $q$ and if there are no constraints on filter latencies and if there is no requirement to predict the state at the time of the next sample (using $q = -1$) then $q$ should be chosen to minimize the variance of a given state estimate, e.g. for the smoother with $k_t = 0$. Expanding (52c) above for the $k_t$th element along the diagonal of **VRF**, i.e. the variance reduction factor of the $k_t$th estimate $\text{VRF}[k_t, k_t]$, yields a polynomial in $q$, i.e. $\mathcal{V}_{k_t}(q)$. Stationary values of $q$ that locally minimize (or maximize) $\mathcal{V}_{k_t}$ are found by solving

$$\dot{\mathcal{V}}_{k_t}(q) = 0, \text{ i.e. the roots of } \dot{\mathcal{V}}_{k_t}(q) \text{ where} \tag{53a}$$

$$\dot{\mathcal{V}}_{k_t}(q) = d\mathcal{V}_{k_t}(q)/dq \,. \tag{53b}$$

Erlang weights for various shape parameters ($\kappa = 0, 1, 2, 4 \,\&\, 8$) are plotted in Figure 7, along with a uniform (i.e. constant) weight with a finite length ($M = 24$). The $p$ parameter of each Erlang weight is set so that its VRF is the same as the VRF of the rectangular weight ($\text{VRF} = 0.04017$) which ensures that all weights attenuate white noise by the same amount, for a fair comparison. To compute the VRF of these weights, a trivial regression model with $K_X = 1 \,\&\, K_t = 1$ are used in (52) above. Note that as the order (i.e. $K = \kappa + 1$) of the Erlang weight increases, in addition to delaying its centroid and improving its symmetry (i.e. reducing the skew), the smoothing parameter (i.e. $p$) decreases, for an increased rate of exponential decay, i.e. 'thinner tails' and a more 'compact' or 'concentrated' response in time. The way in which the weight parameters (i.e. $\kappa \,\&\, p$) shape the response of the regression filter are discussed further in the next section.

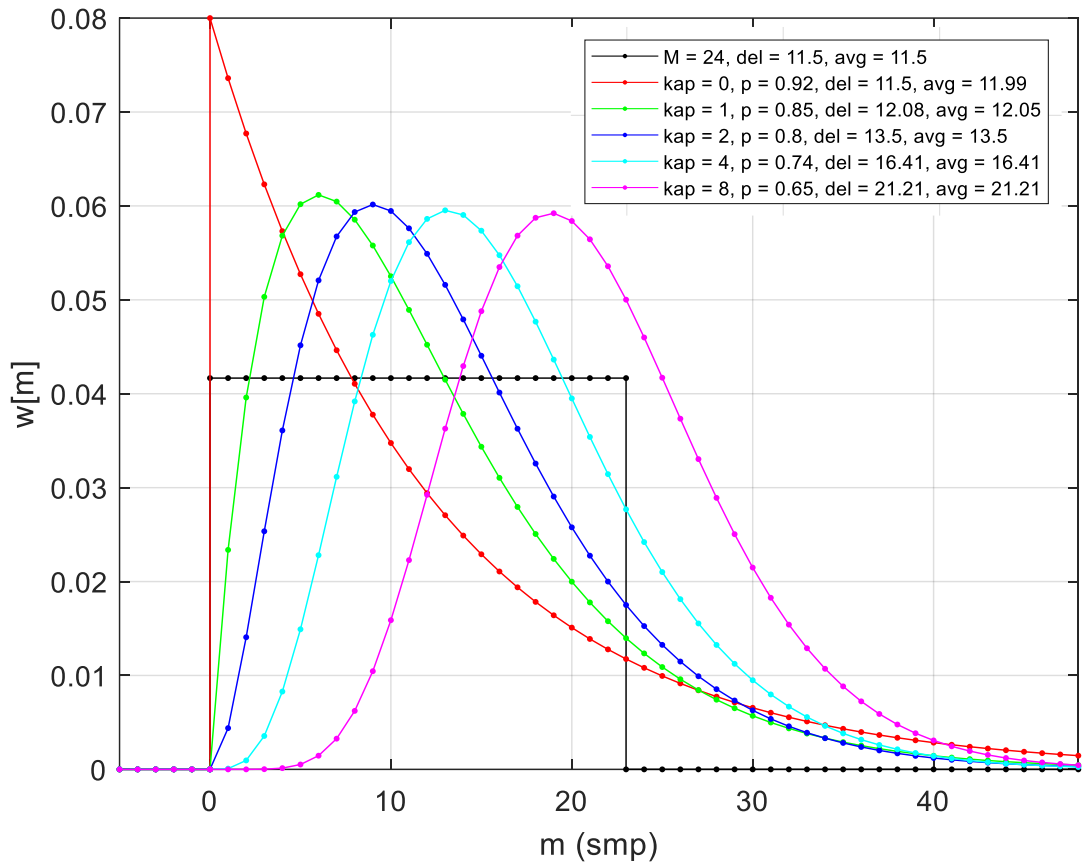

*Figure 7. Normalized Erlang weights for various shape parameters (i.e. κ, coloured lines, see legend) compared with an M-sample rectangular weight (black line). In the Erlang case, the smoothing parameters (i.e. p, see legend) are set so all weights have the same variance reduction factor as the rectangular weight ($VRF = 0.0417$ for $M = 24$). As the shape parameter of the Erlang weight increases, its skew decreases (i.e. it becomes more symmetric) and its lag increases (i.e. it shifts to the right). The dc group-delay (in samples) for the moving averages computed using these weights is also shown (see 'del' in legend). The delay approaches the mean of the continuous Erlang distribution (see 'avg' in legend) as the shape parameter increases.*

## Section 4. Tuning considerations

The term 'bandwidth' is now widely used in technical and non-technical fields alike. It was originally coined by scientists and engineers to describe the range of frequencies that characterise the behaviour of a system or the character of a signal. It soon became synonymous with the capacity of a communication link to transfer data or information. Control engineers see it as the ability of a regulator or servomechanism to quickly (and deftly) adapt to change and now this time-domain interpretation is also widely used in business, management, and the arts.

A state estimator with adequate bandwidth is desirable because unexpected process disturbances, or signals that do not perfectly match the polynomial model, are accommodated. But having too much bandwidth is undesirable because random measurement errors (i.e. white noise and coloured high-frequency interference) are also amplified, which gives rise to 'jittery' or 'hyperactive' systems that react to random fluctuations that are of no material consequence.

Polynomials are a reasonable approximation of very low-frequency signals, and the bandwidth of a polynomial smoother is the frequency at which this approximation is no longer considered to be good enough for a sinusoidal input. Regression weights that are more concentrated in time yield smoothers with greater bandwidth because a low-order polynomial is a reasonable approximation of a high-frequency sinusoid over a sufficiently small timescale. Regression models of high order (i.e $K_X \gg 1$) also yield greater bandwidth because higher frequencies can be approximated (over longer timescales) using high-order polynomial expansions.

Pure integrators and ideal differentiators (in continuous time) represent the two extremes of infinitely narrowband and infinitely wideband systems. They are limiting mathematical cases (their impulse responses are unbounded, and their Fourier transforms are undefined) that are physically unrealizable and ultimately impractical; however, an understanding of these conceptual extremes is helpful when seeking the optimal balance between bandwidth and white-noise gain in a smoother (the $k_t = 0$ case) or differentiator (the $k_t = 1$ case). An integrator (with a Laplace transform of $1/s$) has an infinitely narrow bandwidth in frequency and is infinitely delocalized in time, for infinite attenuation of white noise and infinite amplification of a constant input. A differentiator (with a Laplace transform of $s$) has an infinitely wide bandwidth in frequency and is infinitely localized in time, for infinite amplification of white noise and infinite attenuation of a constant input. Thus, for an effective smoother and/or differentiator we need to increase the maximum frequency (i.e the bandwidth) of the integrator for more concentration in time; and reduce the maximum frequency of the differentiator for less concentration in time.

Polynomial regression provides an intuitive framework for the shaping of a digital filter's response for the desired resolution (for a system) or concentration (for a signal) in time and frequency. In a smoother or differentiator, we want high concentration in time to resolve sudden system disturbances, but we also want high concentration in frequency to attenuate white noise or coloured (high frequency) noise. With appropriately designed filters we can have both but only up to a limit. The extent to which we have both is readily quantified using an uncertainty product $\Delta_t \Delta_\Omega$ where $\Delta_t$ and $\Delta_\Omega$ are arbitrary measures of uncertainty (the inverse of certainty or likelihood concentration) in time and frequency that may be used to characterise the timescale and bandwidth of the analysis. In theory, localization in one domain (i.e. time or frequency) may be over a finite interval with an abrupt or hard cut-off, for a sinc response over an infinite extent in the other domain. Alternatively, localization may be over an infinite interval with a gradual or soft roll-off in both domains. The value of the limit depends on the definition of the uncertainty measure and the achieved product depends on the form of the functions under consideration.

For discounted least-squares regression filters, $\Delta_\Omega$ (i.e. the bandwidth) increases with $K_X$ (i.e. the order of the polynomial model) and with $1/\lambda_w$ (i.e. the rate of the weight's exponential decay or the reciprocal of timescale). It will be shown below that for recursive regression filters with Erlang weights, a given $\Delta_\Omega$ yields a smaller $\Delta_t$ thus the achievable limit is approached as $\kappa$ (i.e. the shape parameter of the regression weight) is increased. This is due to the thinning and tapering of the response tails in both time and frequency. Long tails in an impulse response (in the time domain) are undesirable because very large inputs that are no longer relevant can have an enduring influence on the output. Long tails in a magnitude response (in the frequency domain) are also undesirable because white noise is amplified, and coloured interference has an undue influence on the polynomial fit. Shorter tails in both domains yield more compact responses for improved resolution and a reduced uncertainty product. An absence of infinite tails in one domain leads to sidelobes in the other.

Uncertainty products feature in both engineering and science. For instance, the design and operation of pulse-Doppler tracking radars is governed by products of range & range-rate (i.e. time & frequency) e.g. resolution cells or ambiguity intervals [12]. In quantum mechanics the uncertainty principle is a statement of the wave/particle duality of matter which asserts that

$\sigma_x \sigma_v \geq \hbar$ where (54a)

$\sigma_x$ is the standard deviation of position (54b)

$\sigma_v$ is the standard deviation of momentum and (54c)

$\hbar$ is a fundamental physical constant. (54d)

The $\sigma_x$ and $\sigma_v$ quantities may be interpreted as measures of uncertainty or dispersion (i.e. the reciprocal of certainty, resolution, localization or concentration) thus the state of a system may not be defined precisely and simultaneously in both position and momentum, i.e. uncertainty in one coordinate follows from certainty in the other.

An analogous relationship governs the Fourier analysis and description of linear systems and signals in continuous time [4][5]. For a pulsed signal, or the impulse response of a system, we have the uncertainty or dispersion product:

$\sigma_t \sigma_\Omega \geq 1$ where (55a)

$\sigma_t$ is the standard deviation of the pulse
(i.e. its one-sigma duration) in time (in seconds) and (55b)

$\sigma_\Omega$ is the standard deviation of the pulse
(i.e. its one-sigma bandwidth) in frequency (in radians per second). (55c)

The $\sigma_t$ and $\sigma_\Omega$ metrics are a convenient representation of the $\Delta_t$ and $\Delta_\Omega$ parameters introduced above, i.e. duration in time and bandwidth in frequency, respectively.

Variance in time and frequency (i.e. $\sigma_t^2$ and $\sigma_\Omega^2$) is computed from the zeroth, first, and second moments (i.e. $m_t^0$, $m_t^1$ & $m_t^2$ and $m_\Omega^0$, $m_\Omega^1$ & $m_\Omega^2$) using

$\sigma_t^2 = m_t^2/m_t^0 - (m_t^1/m_t^0)^2$ and (56a)

$\sigma_\Omega^2 = m_\Omega^2/m_\Omega^0 - (m_\Omega^1/m_\Omega^0)^2$ where (56b)

$m_t^k = \int_{-\infty}^{\infty} t^k \psi(t) dt$ and (56c)

$m_\Omega^k = \int_{-\infty}^{\infty} \Omega^k \Psi(\Omega) d\Omega$ with (56d)

$\psi(t)$ being the pulse magnitude (a non-negative function) in the time domain and (56e)

$\Psi(\Omega)$ being the pulse magnitude (a non-negative function) in the frequency domain. (56f)

For the two-sided Gaussian pulse

$\psi(t) = e^{-t^2/2\sigma_t^2}$ and (57a)

$\Psi(\Omega) = \mathcal{F}\{\psi(t)\} = e^{-\sigma_t^2 t^2/2}$ where (57b)

$\mathcal{F}\{\blacksquare\}$ is the Fourier-transform operation over $t = -\infty \ldots \infty$ i.e (57c)

$\Psi(\Omega) = \int_{-\infty}^{\infty} e^{-i\Omega t} \psi(t) dt$ . (57d)

For a one-sided Erlang pulse

$$\psi(t) = t^\kappa e^{-t/\lambda_t} \tag{58a}$$

$$\Psi(s) = \mathcal{L}\{\psi(t)\} = \Gamma(\kappa + 1)/(s + 1/\lambda_t)^{\kappa+1} \text{ then} \tag{58b}$$

$$\Psi(\Omega) = \Psi(s)|_{s=i\Omega} \text{ where} \tag{58c}$$

$\mathcal{L}\{\blacksquare\}$ is the Laplace-transform operation over $t = 0 \ldots \infty$ i.e. (58d)

$$\Psi(s) = \int_0^\infty e^{-st} \psi(t) dt \text{ and} \tag{58e}$$

$\Gamma(\blacksquare)$ is the Gamma function. (58f)

Note that $\sigma_t^2$ and $\sigma_\Omega^2$ are only defined for non-negative functions $\psi(t)$ & $\Psi(\Omega)$ when the transform $\psi(t) \to \Psi(\Omega)$ exists and their respective moments (up to second order) are finite, which is the case for all Gaussian pulses (in both time and frequency) and all Erlang pulses in the time domain. An analytical expression for the variance ($\sigma_t^2$) of an Erlang pulse in continuous time is provided in (8h). However, in the frequency domain, the required moments are only defined for Erlang pulses with $\kappa \geq 3$.

In continuous time the unity limit in (55) is only reached for Gaussians (see Table 3, Table 4 & Table 5). This follows from the remarkable fact that the Fourier transform of a Gaussian pulse in time (centred on $t = 0$) with a standard deviation of $\sigma_t$, is a Gaussian in frequency (centred on $\Omega = 0$) with a standard deviation of $\sigma_\Omega = 1/\sigma_t$. For Erlang pulses, the limit is approached as $\kappa \to \infty$, i.e. as the Erlang pulse becomes more symmetric and more Gaussian-like. This suggests that using $\kappa > 0$ in discounted polynomial regression is desirable because it permits a more focused analysis in both time *and* frequency than would otherwise be achievable using a (one-sided) exponential with $\kappa = 0$. Using a Gaussian weight of infinite extent is ideal in this respect but infeasible. The realization of a Gaussian weight requires an infinite delay to be applied so that the tails of the impulse response extend to positive and negative infinity, thus causal realizations in continuous time are not possible. Erlang weights may be shaped using $\kappa$ to have a finite tail in the forward time direction and an infinite tail in the past, thus causal realizations are feasible in both continuous and discrete time using feedback. The one standard-deviation bandwidth for Erlang windows may also be approximately computed using $\tilde{\sigma}_\Omega \cong \sigma_\Omega = 1/\sigma_t$ with $\sigma_t = T_s\sqrt{\sigma_w^2}$ where $\sigma_w^2$ is evaluated using (8h). The quality of the approximation improves as the Erlang pulse becomes more Gaussian-like (see Table 6).

Table 3. Time dispersion (i.e. $\sigma_t$) of the continuous-time Erlang distribution for various $p$ & $\kappa$ with a standard deviation of $\sigma_t = -1/\ln p$ and a continuous-time Gaussian distribution (right-most column) with the same variance as the exponential.

|  | $\kappa = 0$ | $\kappa = 1$ | $\kappa = 2$ | $\kappa = 3$ | $\kappa = 4$ | $\kappa = 8$ | $\kappa = 12$ | $\kappa = 16$ | $\kappa = 24$ | Gsn |
|---|---|---|---|---|---|---|---|---|---|---|
| $p = 0.7$ | 2.804 | 3.965 | 4.856 | 5.607 | 6.269 | 8.411 | 10.109 | 11.560 | 14.018 | 2.804 |
| $p = 0.8$ | 4.481 | 6.338 | 7.762 | 8.963 | 10.021 | 13.444 | 16.158 | 18.477 | 22.407 | 4.481 |
| $p = 0.9$ | 9.491 | 13.423 | 16.439 | 18.982 | 21.223 | 28.474 | 34.221 | 39.133 | 47.456 | 9.491 |

Table 4. Frequency dispersion (i.e. $\sigma_\Omega$) of the continuous-time Erlang distribution for various $p$ & $\kappa$ and a continuous-time Gaussian distribution (right-most column) with a standard deviation of $\sigma_\Omega = 1/\sigma_t$.

|  | $\kappa = 0$[a] | $\kappa = 1$[b] | $\kappa = 2$[c] | $\kappa = 3$ | $\kappa = 4$ | $\kappa = 8$ | $\kappa = 12$ | $\kappa = 16$ | $\kappa = 24$ | Gsn |
|---|---|---|---|---|---|---|---|---|---|---|
| $p = 0.7$ | - | - | - | 0.357 | 0.252 | 0.146 | 0.113 | 0.095 | 0.076 | 0.357 |
| $p = 0.8$ | - | - | - | 0.223 | 0.158 | 0.091 | 0.071 | 0.060 | 0.048 | 0.223 |
| $p = 0.9$ | - | - | - | 0.105 | 0.075 | 0.043 | 0.033 | 0.028 | 0.023 | 0.105 |

[a] 0th, 1st & 2nd moments are undefined; [b] 1st & 2nd moments are undefined; [c] 2nd moment is undefined.

Table 5. Time-frequency dispersion product (i.e. $\sigma_t \sigma_\Omega$) of the continuous-time Erlang distribution for various $p$ & $\kappa$ and a continuous-time Gaussian distribution (right-most column) with $\sigma_\Omega \sigma_t = 1$.

|  | $\kappa = 0$[a] | $\kappa = 1$[b] | $\kappa = 2$[c] | $\kappa = 3$ | $\kappa = 4$ | $\kappa = 8$ | $\kappa = 12$ | $\kappa = 16$ | $\kappa = 24$ | Gsn |
|---|---|---|---|---|---|---|---|---|---|---|
| $p = 0.7$ | - | - | - | 2.000 | 1.581 | 1.225 | 1.140 | 1.102 | 1.066 | 1.000 |
| $p = 0.8$ | - | - | - | 2.000 | 1.581 | 1.225 | 1.140 | 1.102 | 1.066 | 1.000 |
| $p = 0.9$ | - | - | - | 2.000 | 1.581 | 1.225 | 1.140 | 1.102 | 1.066 | 1.000 |

[a] 0th, 1st & 2nd moments are undefined; [b] 1st & 2nd moments are undefined; [c] 2nd moment is undefined.

Table 6. Approximate frequency dispersion (i.e. $\tilde{\sigma}_\Omega$) of the continuous-time Erlang distribution for various $p$ & $\kappa$ (computed using $\tilde{\sigma}_\Omega = 1/T_s\sqrt{\sigma_w^2}$) and a continuous-time Gaussian distribution (right-most column).

|  | $\kappa = 0$ | $\kappa = 1$ | $\kappa = 2$ | $\kappa = 3$ | $\kappa = 4$ | $\kappa = 8$ | $\kappa = 12$ | $\kappa = 16$ | $\kappa = 24$ | Gsn |
|---|---|---|---|---|---|---|---|---|---|---|
| $p = 0.7$ | 0.357 | 0.252 | 0.206 | 0.178 | 0.160 | 0.119 | 0.099 | 0.087 | 0.071 | 0.357 |
| $p = 0.8$ | 0.223 | 0.158 | 0.129 | 0.112 | 0.100 | 0.074 | 0.062 | 0.054 | 0.045 | 0.223 |
| $p = 0.9$ | 0.105 | 0.075 | 0.061 | 0.053 | 0.047 | 0.035 | 0.029 | 0.026 | 0.021 | 0.105 |

Discrete-time Gaussian filters with a finite impulse response *are* realizable but the tails in both time directions are truncated, which creates sidelobes in frequency, thus the response is no longer Gaussian in frequency or time. If the truncation is too abrupt or if the sampling rate is too low (for efficiency) then there is no guarantee that the discretised Gaussian will be close to the continuous-time limit and other shapes (e.g. Slepian) may yield a lower uncertainty product in non-recursive finite-impulse-response filters.

The uncertainty analysis above is for the Erlang regression weight in isolation. It may be convolved with an input signal to compute its moving average (see Section 1) i.e. for a regressor with $K_X = 1$ and $K_t = 1$. The frequency response of these Erlang-weighted moving averages (with the desired finite bandwidth) are convolved with the frequency responses of the regressors (with an infinitesimal bandwidth) to yield the frequency response of the realized regression filter.

An alternative definition of bandwidth is useful when more general cases with $K_X \geq 1$ and $K_t \geq 1$ are considered (see Section 2). The $K_X = 1$ and $K_t = 1$ case is continued below, then smoothers with $K_X > 1$ and $K_t = 1$ are analysed in the remainder of this section. Using $K_X > K_t$ (i.e. more monomials than derivatives) may be used to increase the bandwidth of the smoother without increasing the white-noise gain if $\kappa > 0$ is also used. More general $K_X > 1$ and $K_t > 1$ filters, i.e. first and second derivatives for gradient and curvature or velocity and acceleration estimates, are applied in the next section.

Even in technical fields there is no universal mathematical definition of bandwidth for a low-pass system. It is usually understood to be the frequency at which the magnitude or the squared magnitude is halved, or twice this frequency. It may also be the frequency (or twice the frequency) at which these quantities become zero, negligible, or one or more standard deviations from dc, or where distortion becomes significant. One standard deviation of the magnitude was used above to quantify impulse-response duration, frequency-response bandwidth, and the way in which the shape parameter of the regression weight focuses the analysis in both time and frequency. An alternative definition will now be used, to analyse the factors that determine the bandwidth and white-noise gain (i.e. the variance reduction factor) of the resulting regression filter, i.e. a smoother, differentiator, or (derivative) state estimator.

The desired magnitude of a smoothing filter, i.e. a regression filter with $k_t = 0$, is unity at dc, where $\omega = 0$. Note that for a dc signal, or a constant offset of infinite extent, phase has no meaning. As discussed at length above, we also require a non-negligible bandwidth, thus unity magnitude near dc is also desirable, where 'near' is a deliberately vague term as these smoothers are not designed using a cut-off frequency specification. The bandwidth is instead set from the desired concentration in time (i.e. the regression weight shape and extent, as determined using $\kappa$ & $p$) and the regression model 'flexibility' (i.e. the polynomial degree which is equal to $K_X - 1$).

Over the near-dc passband the desired magnitude is unity and the desired group-delay is $q$. Thus, the desired low-frequency response of the smoother is

$$D(\omega) = e^{-iq\omega}. \tag{59a}$$

A response of $D(\omega) = 0$ is desired in the far-from-dc stopband. A gradual magnitude roll-off or a wide transition band in between these pass- and stop- 'bands' is acceptable as this yields a Gaussian-like magnitude response for a more compact Gaussian-like impulse response in the time domain.

The complex response error is the difference between the desired response and the realized response, i.e.

$$E(\omega) = H(\omega) - D(\omega) \text{ where} \tag{60a}$$

$H(\omega)$ is the frequency response of the smoother.

The frequency at which the squared magnitude of the response error, i.e. $|E(\omega)|^2$ or the 'distortion', is equal to $1/2$, i.e. the 3-dB point, is a convenient measure of a regression smoother's bandwidth. This frequency ($f_c$ in cycles per sample or $\omega_c$ in radians per sample) is referred to here as the distortion-free bandwidth. The magnitude error and the phase error i.e.

$$|H(\omega)| - |D(\omega)| \text{ and} \tag{61a}$$

$$\angle H(\omega) - \angle D(\omega) \tag{61b}$$

are usually used to assess the response of recursive IIR filters; however, these functions roll-off at different rates for different filters, thus when magnitude flatness and phase linearity are equally important, the combined error metric in (60) is a better indication of a recursive smoother's bandwidth.

Increasing the order of the polynomial model (i.e. $K_X$) increases the distortion-free bandwidth of the smoother because it ensures that the first $K_X$ derivatives of the smoother's frequency response $H(\omega)$ and the desired frequency response $D(\omega)$ are the same, i.e.

$$\left\{\frac{d^{k_\omega}}{d\omega^{k_\omega}}H(\omega)\right\}\bigg|_{\omega=0} = \left\{\frac{d^{k_\omega}}{d\omega^{k_\omega}}D(\omega)\right\}\bigg|_{\omega=0} = (-iq)^{k_\omega} \text{ for } k_\omega = 0 \ldots K_X - 1 \,. \tag{62a}$$

The delay parameter ($q$) sets the time-shift applied for a 'truly polynomial' signal. For other 'almost polynomial' signals, e.g. low-frequency sinusoids, the applied phase-delay is only approximately equal to $q$. The distortion-free bandwidth (or the flatness of the magnitude response and the linearity of the phase response) indicates whether low-frequency sinusoids are 'sufficiently polynomial' for reasonable estimates using a given regression filter.

The magnitude and phase response of some Erlang weights and the rectangular weight in Figure 7 is shown in Figure 8, along with the low-frequency magnitude and phase errors. The ideal phase response for the delay applied by each window is shown using the dotted lines. Ideally the phase shift (in radians) applied by the window should be proportional to the angular frequency (in radians per sample) with the constant of proportionality equal to $-q$ (in samples) for a linear-phase response; however, the deviation from linearity increases with frequency for the Erlang weights. The low-frequency response distortion, i.e. $|E(\omega)|^2$, for the Erlang weights in Figure 7 & Figure 8 is plotted in Figure 9 (thick coloured lines) along with the magnitude error (thin coloured lines). For Erlang weights with $p$ set for the same VRF, these plots indicate that larger shape parameters (i.e. $\kappa$): improve phase linearity, increase the width of the passband (for reduced distortion of low-frequency signals) and decrease gain in the stopband (for increased attenuation of high-frequency noise).

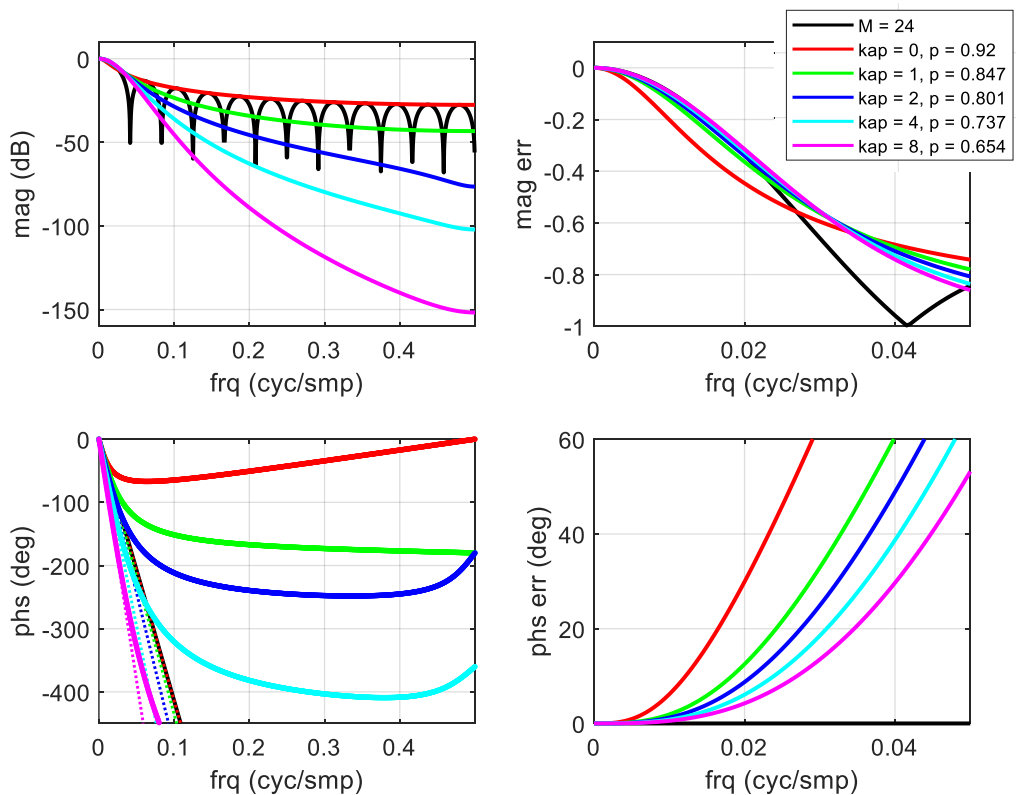

*Figure 8. Frequency responses of the weights in Figure 7. Magnitude response (decibel scale, top left), phase response (degrees, bottom left), magnitude error (linear scale, top right), phase error (degrees, bottom left) as a function of relative frequency (f in cycles per sample). The full Nyquist band is shown on the left; the low-frequency magnitude and phase errors are shown on the right. As the shape parameter (i.e. κ) of the Erlang window increases, its high-frequency magnitude attenuation increases (see upper subplots) and its low-frequency phase-linearity improves (see lower subplots).*

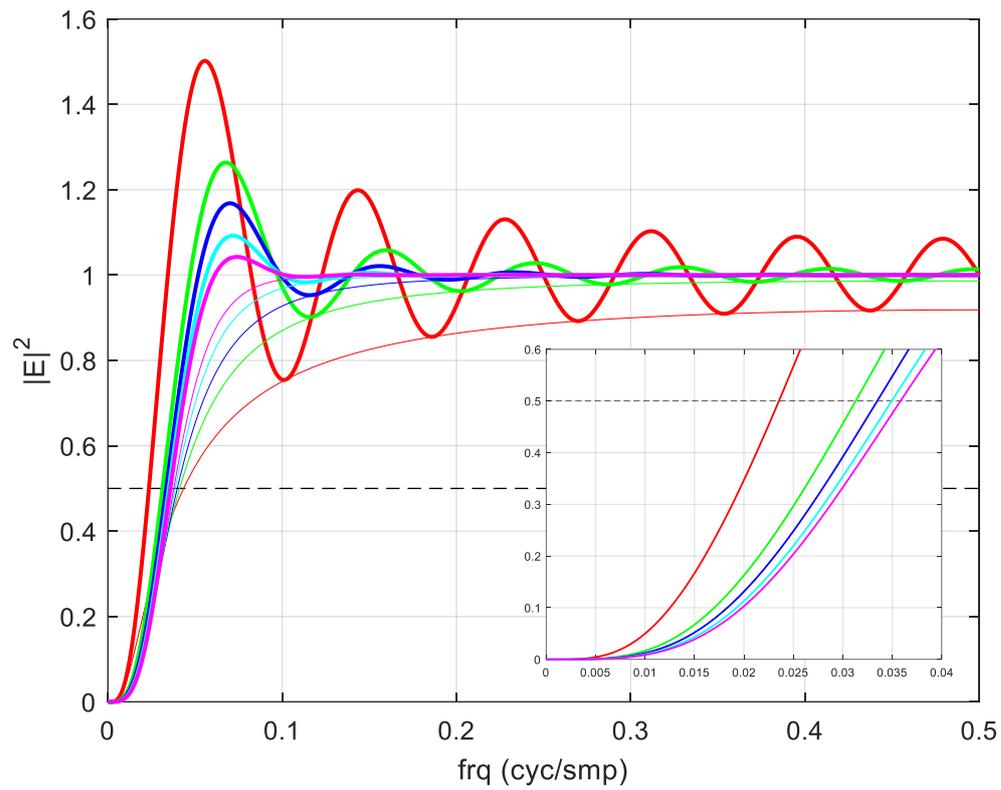

*Figure 9. Frequency response distortion for the Erlang weights in Figure 7 & Figure 8. Inset shows the low-frequency region. The low-frequency distortion (i.e. phase linearity and magnitude flatness) decreases as the shape parameter increases and the 3 dB distortion-free bandwidth increases (see inset).*

The optimal delay ($q$) that minimizes the $k_t = 0$ element of the VRF matrix diagonal, the resulting VRF, and the distortion-free bandwidth ($f_c$), of smoothing filters for various $K_X$, $\kappa$ & $p$ combinations are shown in Table 7 & Table 8. Their corresponding frequency responses are plotted in Figure 10 & Figure 12 along with their frequency response distortion in Figure 11 & Figure 13. These results illustrate some of the relationships between the various parameters and the response of the resulting smoother that have been discussed above. In particular, the following observations are worth noting:

- For $p = 0.8$, using $K_X = 3$ & $\kappa = 0$ (dashed red line) yields a wider bandwidth but a higher variance than using $K_X = 2$ & $\kappa = 0$ (solid red line); however, using $K_X = 3$ & $\kappa = 2$ (dashed blue line) yields a wider bandwidth *and* a lower variance (see Table 7, Figure 10 & Figure 11).
- For $K_X = 2$, using $\kappa = 0$ & $p = 0.80$ (solid green line) yields a wider bandwidth but a higher variance than using $\kappa = 0$ & $p = 0.85$ (solid blue line); however, using $\kappa = 3$ & $p = 0.75$ (dashed red line) yields a wider bandwidth *and* a lower variance (see Table 8, Figure 12 & Figure 13).

Note also that these improvements in bandwidth and variance, resulting from non-zero shape parameters (i.e. $\kappa$), are only realized if the applied delay (i.e. $q$) is equal to the optimal value (that minimizes the VRF) which increases with the shape parameter (see Table 7 & Table 8). For all weights and all model orders the optimal delay is close to the centroid of the weight, where there is ample data support. Any delay may be applied by setting the delay parameter arbitrarily (e.g. to meet latency requirements); however, the noise gain (i.e. VRF) increases and the bandwidth (i.e. $f_c$) decreases, if the polynomial fit (or its derivatives) are evaluated too far from the centroid of the weight, where data support is diminished (e.g. into the future).

Table 7. Optimal delay, VRF and bandwidth of smoothing filter for various $K_X$ & $\kappa$ combinations.

| $K_X$ | 2 | 2 | 2 | 3 | 3 | 3 |
|---|---|---|---|---|---|---|
| $\kappa$ | 0 | 1 | 2 | 0 | 1 | 2 |
| $p$ | 0.80 | 0.80 | 0.80 | 0.80 | 0.80 | 0.80 |
| $q$ | 8.50 | 13.50 | 17.93 | 5.22 | 9.03 | 12.39 |
| VRF[0,0] | 0.056 | 0.042 | 0.035 | 0.083 | 0.063 | 0.052 |
| $f_c$ | 0.042 | 0.034 | 0.029 | 0.068 | 0.052 | 0.043 |

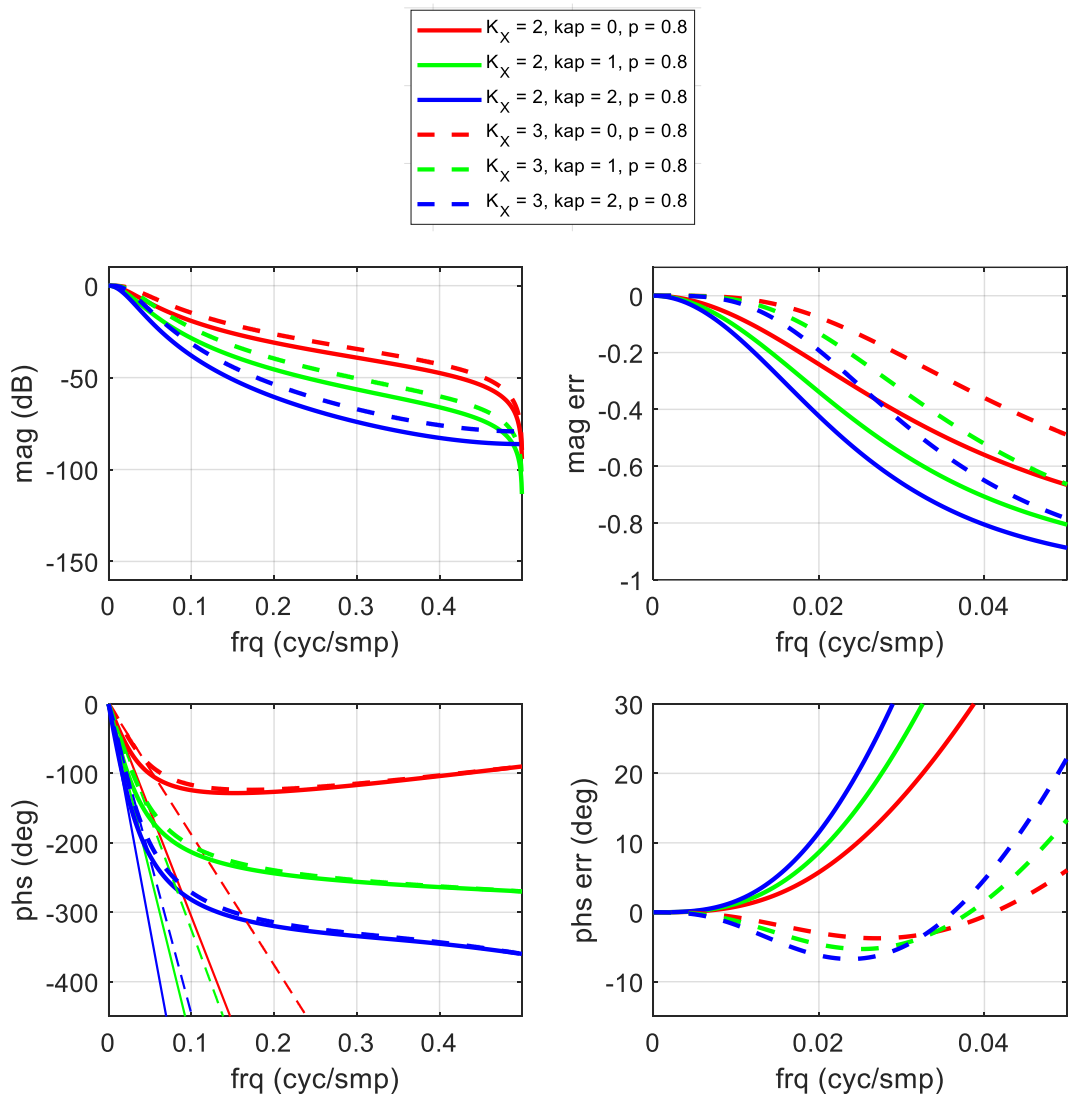

Figure 10. Frequency response of smoothing filter for various $K_X$ & $\kappa$ combinations.

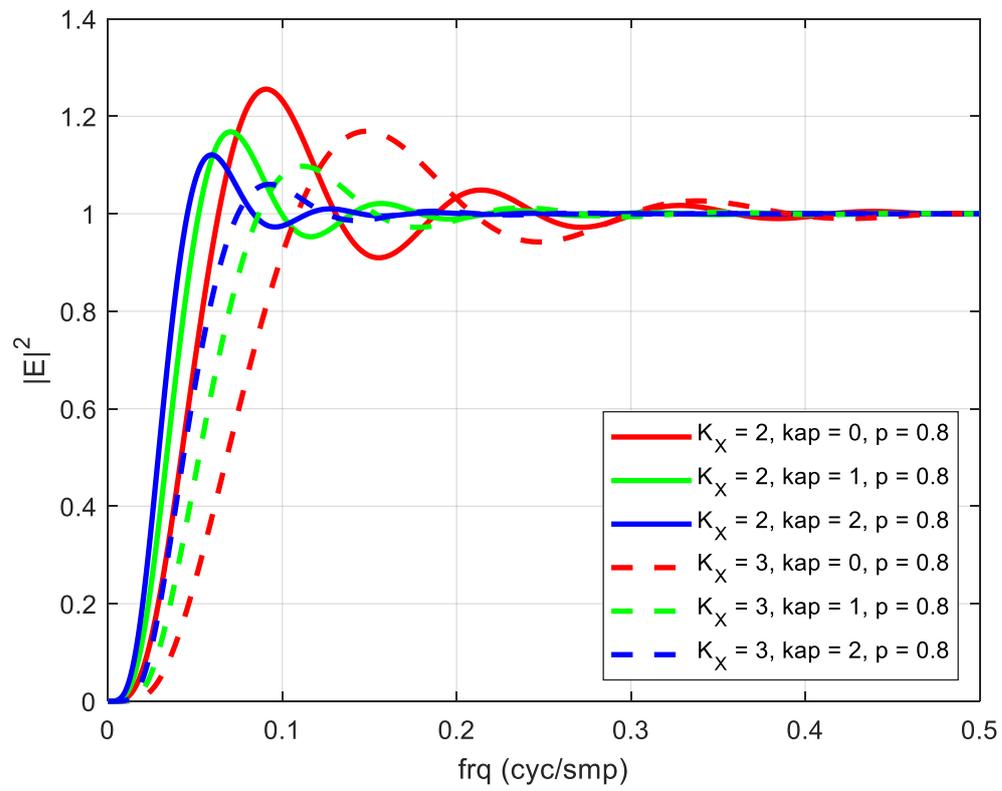

*Figure 11. Frequency response error (squared) of smoothing filter for various $K_X$ & $\kappa$ combinations.*

Table 8. Optimal delay, VRF and bandwidth of smoothing filter for various $\kappa$ & $p$ combinations.

| $K_X$ | 2 | 2 | 2 | 2 | 2 | 2 |
|---|---|---|---|---|---|---|
| $\kappa$ | 0 | 0 | 0 | 3 | 3 | 3 |
| $p$ | 0.75 | 0.80 | 0.85 | 0.75 | 0.80 | 0.85 |
| $q$ | 6.50 | 8.50 | 11.83 | 17.38 | 22.41 | 30.77 |
| VRF[0,0] | 0.071 | 0.056 | 0.041 | 0.039 | 0.031 | 0.022 |
| $f_c$ | 0.054 | 0.042 | 0.031 | 0.033 | 0.026 | 0.019 |

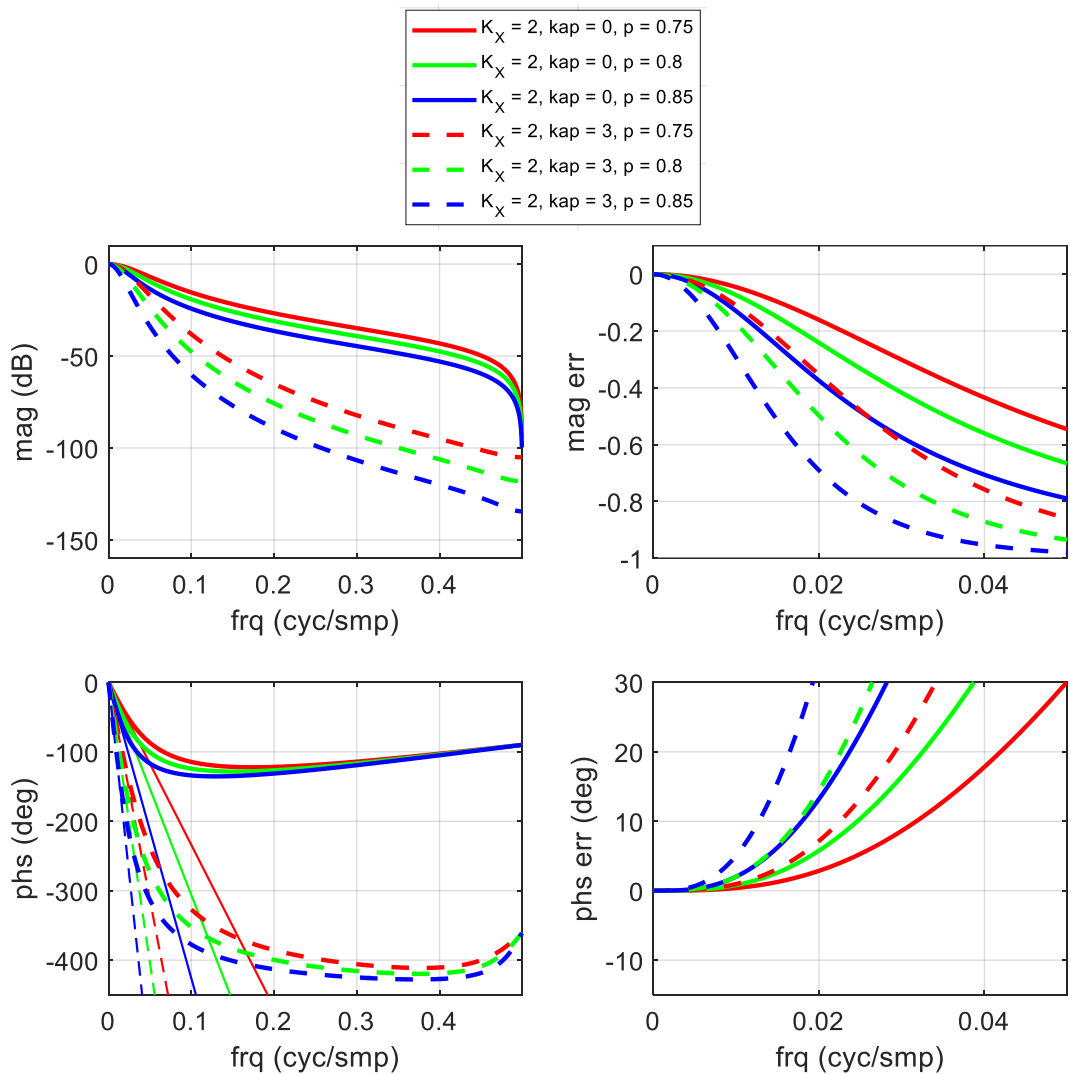

Figure 12. Frequency response of smoothing filter for various $\kappa$ & $p$ combinations.

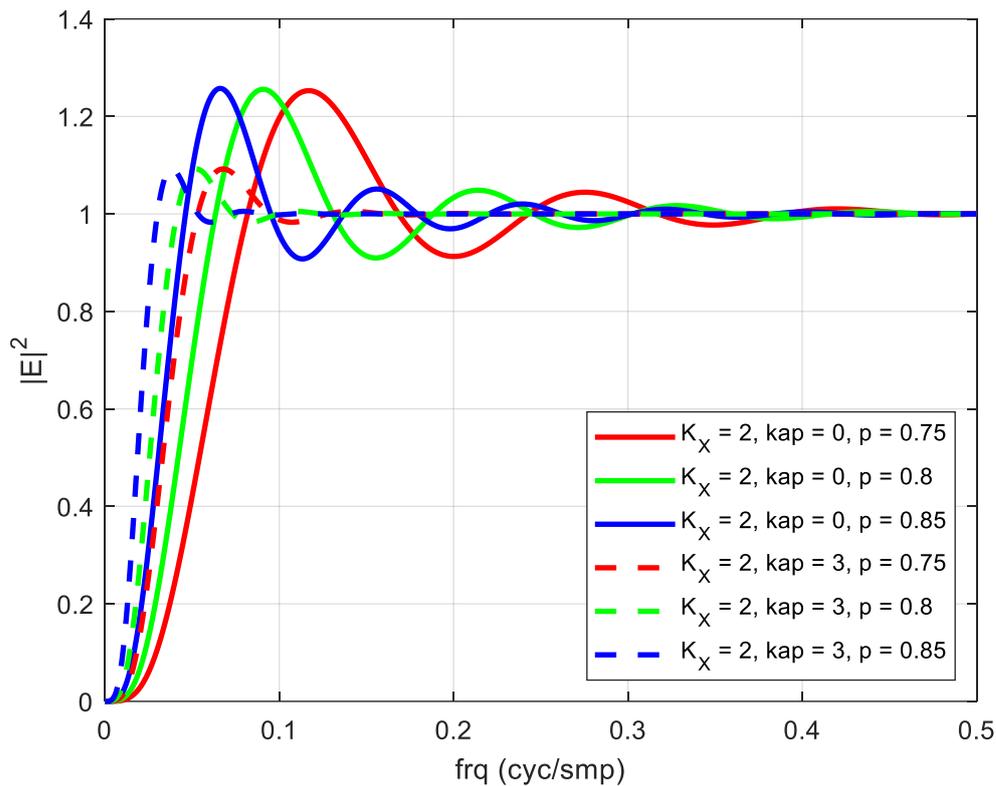

*Figure 13. Frequency response error (squared) of smoothing filter for various κ & p combinations.*

## Section 5. Use cases

In the context of tracking, smoothing, and state estimation, the bandwidth of a filter or 'observer' determines its ability to accommodate unmodelled system behaviour, by removing bias errors. Unfortunately, the cost of higher bandwidth is higher white-noise amplification, which yields larger random errors, for instance, due to uncorrelated measurement noise; thus, there is a fundamental trade-off to consider and some design issues to ponder. For instance: Is the model sufficient and what is the expected frequency, intensity, form, and duration, of model departures and disturbance inputs?

Randomly occurring Gaussian-distributed system disturbances (i.e. process noise) are a convenient assumption and when system observations are degraded by Gaussian-distributed random errors (i.e. measurement noise) the Kalman filter yields the optimal gain [2][8][9]. Gaussian disturbances are a reasonable approximation in many cases, e.g. the motion a helium balloon aloft; however, they are a poor approximation for volition-driven systems, e.g. the motion of a pedestrian. Process noise and measurement noise may be random and uncorrelated (i.e. white); however, Gaussian distributions are a poor approximation for events that are rare but extreme, i.e. distributions with 'fat' or 'heavy' tails. For the regression filters considered here, disturbance inputs do not conform to any known distribution; furthermore, measurement noise is not assumed to be Gaussian (although it is assumed to be white) and its variance is unknown.

A wideband (open-loop) state estimator and a (closed-loop) feedback controller are more able to track unknown process inputs (random or deterministic) that are more localised in time (i.e. sudden or abrupt), where 'tracking' is the ability of the estimator to follow the true state of the system without bias error or random error [11]. Bias errors are temporally correlated, i.e. they are not 'white'; furthermore, they may be transient (e.g. in response to a sudden change of state caused by an unknown disturbance input) or enduring (e.g. due to process model errors or approximations). Unfortunately, white noise (distributed uniformly over the full Nyquist band, i.e. from zero to half the sampling frequency) is amplified as the bandwidth of a filter expands and a point is eventually reached where its response in frequency has uniform magnitude and its response in time is a delayed impulse; thus it does no smoothing or state estimation. Regression provides an intuitive framework for balancing bias and random errors when disturbance inputs are unknown and measurement errors are white and the production of a covariance matrix for state estimates is a bonus.

Digital linear-phase FIR realizations of polynomial regression filters are sometimes known as Savitzky-Golay filters – named after the physical chemists who first applied them to the smoothing of spectra. They are commonly used in: image-processing [17][18][19], power-engineering[20][21] and bio-medical applications [22][23][24][25][26]. In addition to data-smoothing applications, recursive realizations of FIR Savitzky-Golay filters have also been used to extract features for the online segmentation of noisy time-series data [14]. In target-tracking applications, polynomial regression filters with an IIR, that are designed using an exponential error weight, are sometimes used instead of Kalman filters [1][2][8][13][27]. IIR realizations with Erlang weights are used in all the use cases considered in this section. The MATLAB® code used to design the filters and generate the results is available from the author's research website [28].

### Target tracker

Consider a massive object that moves in free space according to Newton's laws of motion, i.e. an integrating system with two poles in the complex plane at $s = 0$. The object (i.e. the 'target') is observed using an imaging sensor (e.g. an infra-red video camera) and its motion in one dimension (e.g. in the vertical coordinate of the image plane) may be modelled using the following linear discrete-time state-space system:

$$\boldsymbol{w}[n] = \boldsymbol{G}\boldsymbol{w}[n-1] + \boldsymbol{H}x[n] \text{ and} \tag{63a}$$

$$y[n] = \boldsymbol{C}\boldsymbol{w}[n] \text{ where} \tag{63b}$$

$$\boldsymbol{H} = \begin{bmatrix} \frac{1}{2}T_s^2 \\ T_s \end{bmatrix}, \boldsymbol{G} = \begin{bmatrix} 1 & T_s \\ 0 & 1 \end{bmatrix} \text{ and } \boldsymbol{C} = \begin{bmatrix} 1 & 0 \end{bmatrix}. \tag{63c}$$

The scalar input $x[n]$ is the acceleration component resulting from a net force (e.g. from a jet engine) that is held constant between samples, i.e. over one full sample period. The $k = 0$ and $k = 1$ elements of the internal state vector $\boldsymbol{w}[n]$ are the position and velocity components of the object (respectively) in the given dimension and the output $y[n]$ of the system is the position coordinate. The sensor 'observes' the noisy sequence of measurements $y[n] + \varepsilon[n]$, where $\varepsilon[n]$ is the output of a white-noise process with a mean of zero and a variance of $\sigma_\varepsilon^2$. A synthetic measurement sequence ($N = 200$) was generated using $\sigma_\varepsilon = 0.5$ (pixels or 'pix'), $T_s = 0.01$ (seconds or 'sec'), i.e. for a video camera with a frame-rate of 100 Hz. Two scenarios were considered.

In the first scenario (see Figure 14 to Figure 17) the system is driven by a *constant* acceleration input, of $x[n] = -20.0$ (pix/sec/sec), for $n = 0 \dots N - 1$ from its initial state of $w_0[0] = 1000.0$ (pix) & $w_1[0] = 20.0$ (pix/sec) at $n = 0$. The output of this system, for a constant acceleration input, in the absence of measurement noise, is a quadratic curve; therefore, we choose a model with three monomial components (i.e. $K_X = 3$). We would also like to estimate the position & velocity states, and the acceleration input; therefore, we estimate the first three temporal derivatives (i.e. $K_t = 3$). Moderate lags are assumed to be acceptable thus we use an Erlang weight with $\kappa = 2$ for improved low-frequency (phase and magnitude) linearity and white-noise attenuation. Erlang weights with $p =$

0.6 & 0.9 are considered for state estimators with a wide (Figure 14 & Figure 15) and narrow (Figure 16 & Figure 17) bandwidth, respectively; see Table 9 for filter parameters.

Outputs of the state estimator are shown in Figure 15 & Figure 17. Estimates of target position ($k_t = 0$) velocity ($k_t = 1$) and acceleration ($k_t = 2$) are plotted (blue lines) along with their 3-sigma confidence limits, $\pm 3\hat{\sigma}_{k_t}$ (magenta lines); these limits are computed using the estimate of the measurement-noise variance $\hat{\sigma}_\varepsilon^2$ using (35) with the recursions in (38)-(41); the $\pm 3\hat{\sigma}_\varepsilon$ limits (green lines) are shown in the top subplot of each figure. The true position, velocity and acceleration of the target are also plotted (red lines). In the absence of bias errors, target measurements (greed dots) outside the $\pm 3\hat{\sigma}_\varepsilon$ limits should be scarce and excursions of the true target states beyond the $\pm 3\hat{\sigma}_{k_t}$ limits should be rare, after the initiation transient has passed and in the absence of manoeuvres. The limits could be used as 'gates' for data association and $1/\hat{\sigma}_\varepsilon^2$ could be used as a measure of confidence for track management; however, these practical matters are not considered here. The outputs of the state estimator are advanced in time, i.e. shifted to the left, by $q$ samples in these plots to compensate for the filter group delay, so that the estimates are aligned with the truth in time, to assist with visualization and (qualitative) accuracy assessment. For the $p = 0.6$ & $0.9$ filters, $q = 5.41$ & $26.23$, respectively (see Table 9).

The 'internal workings' of the state estimator at an arbitrary moment in time ($n = 155$) are shown in Figure 14 & Figure 16. The regression weight is plotted in the lower subplot (cyan line). The resulting polynomial fit is shown in the upper subplot (blue line) along with the 3-sigma limits (magenta lines) and the true position of the target (red line). The output of the smoother at that moment (i.e. the position estimate) is computed by evaluating the fitted polynomial at $n - q$ (blue dot and circle). Velocity and acceleration estimates are computed by evaluating the first and second derivatives of the polynomial at that time (not shown). Note that for the zeroth derivative (i.e. for $k_t = 0$) of a third-order polynomial (i.e. a quadratic curve with $K_X = 3$) the 3-sigma limits have two stationary points: two local minima and one local maximum. During the design process, the delay parameter of the state estimator (i.e. $q$) is chosen to be at the stationary point with the least delay, where the variance reduction factor of the position estimate is a minimum, using (52) & (53).

Figure 14 suggests that the bandwidth (thus the white-noise gain) of the state estimator is too large. For a quadratic fit, the duration of the regression weight is insufficient to attenuate the measurement noise (i.e. the VRF is too high) thus incorrect inferences are drawn, concerning the target's position in the distant past and future. Gross errors are observed over time; however, the estimator appears to be 'aware' of the problem as the variance grows rapidly in both directions. Although, the position estimate at the evaluation point is reasonable. Figure 15 indeed shows that the position estimate is acceptable over the entire target's trajectory; however, the velocity estimate is overly noisy, and the acceleration estimate is meaningless. The situation is greatly improved when a regression weight with a longer memory is used for a narrower bandwidth (see Figure 16 & Figure 17). The motion of the target is now clearly resolved (eventually) at the cost of a greater lag and a longer initiation transient.

In the second scenario (see Figure 18, Figure 19 & Figure 20) the constant acceleration input is replaced by a *random* acceleration input, that is drawn from a Gaussian distribution with a mean of zero and a standard deviation of 100 (pix/sec/sec). The state of the system, i.e. the Newtonian motion of the target, is constantly perturbed by the process noise input, thus there is no point trying to estimate the target's acceleration, therefore $K_t = 2$ is used. For derivative estimates, $K_X \geq K_t$ is required; although $K_X = 2$ is used in this used case to reduce the VRF and filter complexity. The outputs of state estimators with $\kappa = 0$ & $p = 0.85$ (for high gain) $\kappa = 2$ & $p = 0.85$ (for low gain) and

$\kappa = 3$ & $p = 0.75$ (for medium gain) are shown in Figure 18, Figure 19 & Figure 20, respectively; filter parameters are shown in Table 9.

The first estimator with an exponential weight ($\kappa = 0$) has a large random error, particularly in the velocity estimate. The output of the second estimator with an increased shape parameter ($\kappa = 2$) is over smoothed, resulting in persistent velocity excursions, which yield large bias errors in the position; for instance, the truth lingers outside the 3-sigma limits of the position estimate around $n = 120$ for many updates. Random errors and bias errors are well balanced in the third estimator with an increased shape parameter ($\kappa = 3$) and a decreased smoothing parameter. The smoothed output (i.e. the $k_t = 0$ state) of this medium-gain estimator has a wider bandwidth ($f_c$) *and* a lower variance (VRF[0,0]) than the high-gain estimator (see Table 9), thus it tracks the randomly perturbed target with reduced bias *and* less jitter.

*Table 9. Parameters of regression filters used in target tracking use-case*

| $K_X$ | 3 | 3 | 2 | 2 | 2 |
|---|---|---|---|---|---|
| $\kappa$ | 2 | 2 | 0 | 2 | 3 |
| $p$ | 0.60 | 0.90 | 0.85 | 0.85 | 0.75 |
| $q$ | 5.41 | 26.23 | 11.83 | 24.61 | 17.38 |
| VRF[0,0] | 0.1197 | 0.0247 | 0.0405 | 0.0254 | 0.0393 |
| $f_c$ | 0.0989 | 0.0205 | 0.0306 | 0.0210 | 0.0330 |

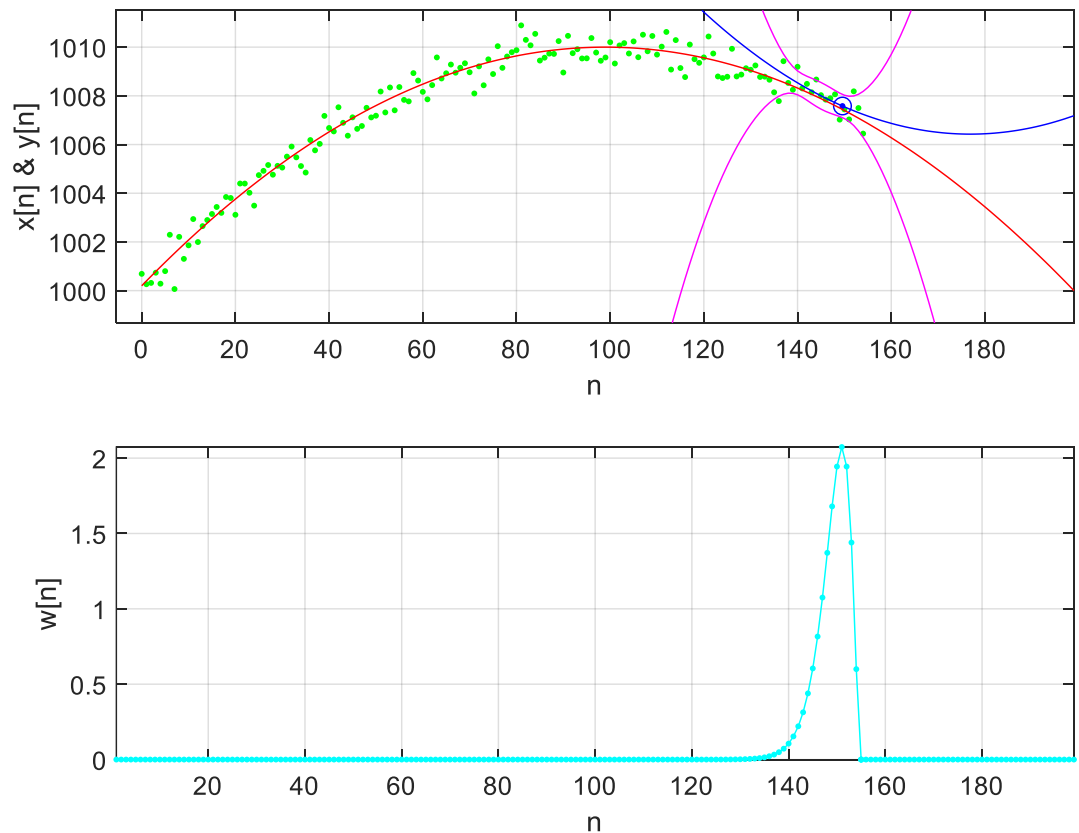

*Figure 14. Constant-acceleration tracking scenario. High-gain state estimator with $K_X = 3$, $\kappa = 2$ & $p = 0.6$. Regression weight on lower subplot; weighted least-squares polynomial fit in upper subplot, with true target position (red), position estimate (blue) 3-sigma limits (magenta) and noisy measurements (green).*

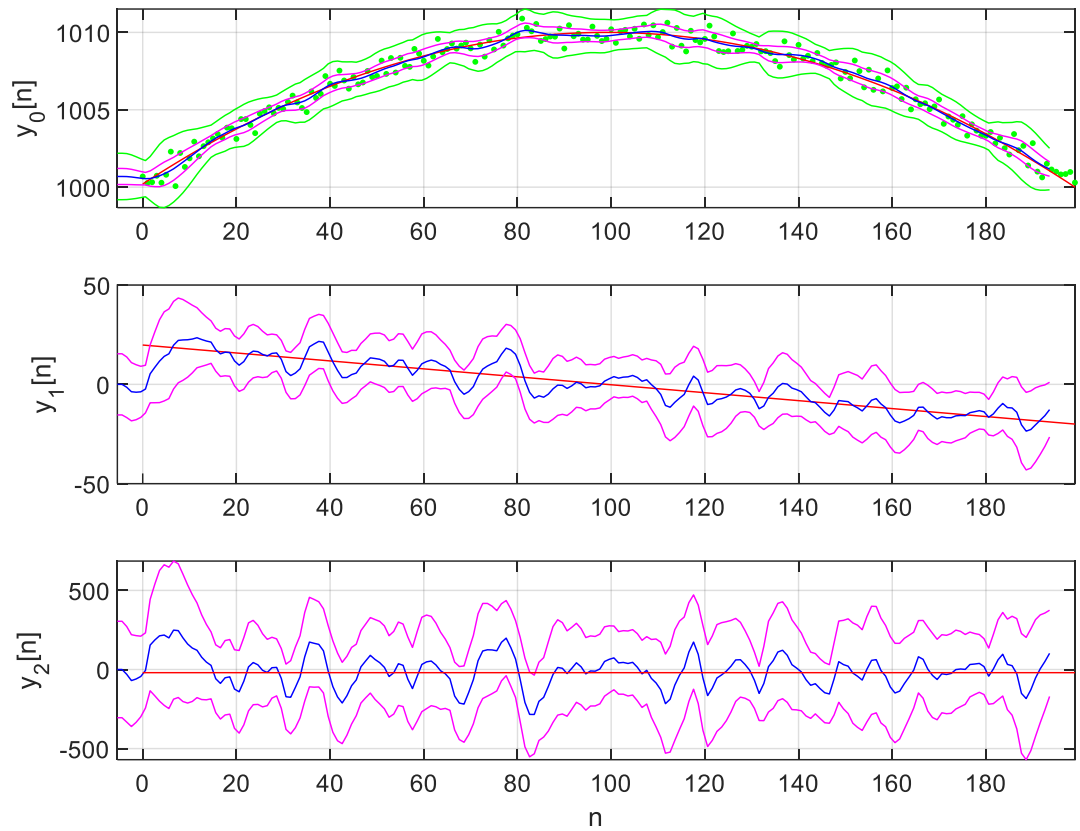

Figure 15. Constant-acceleration tracking scenario. Output of high-gain state estimator with $K_t = 3$, $K_X = 3$, $\kappa = 2$ & $p = 0.6$. Position (with measurements in green), velocity and acceleration estimates and 3-sigma limits (blue and magenta lines) shown in the upper, middle and lower subplots, respectively.

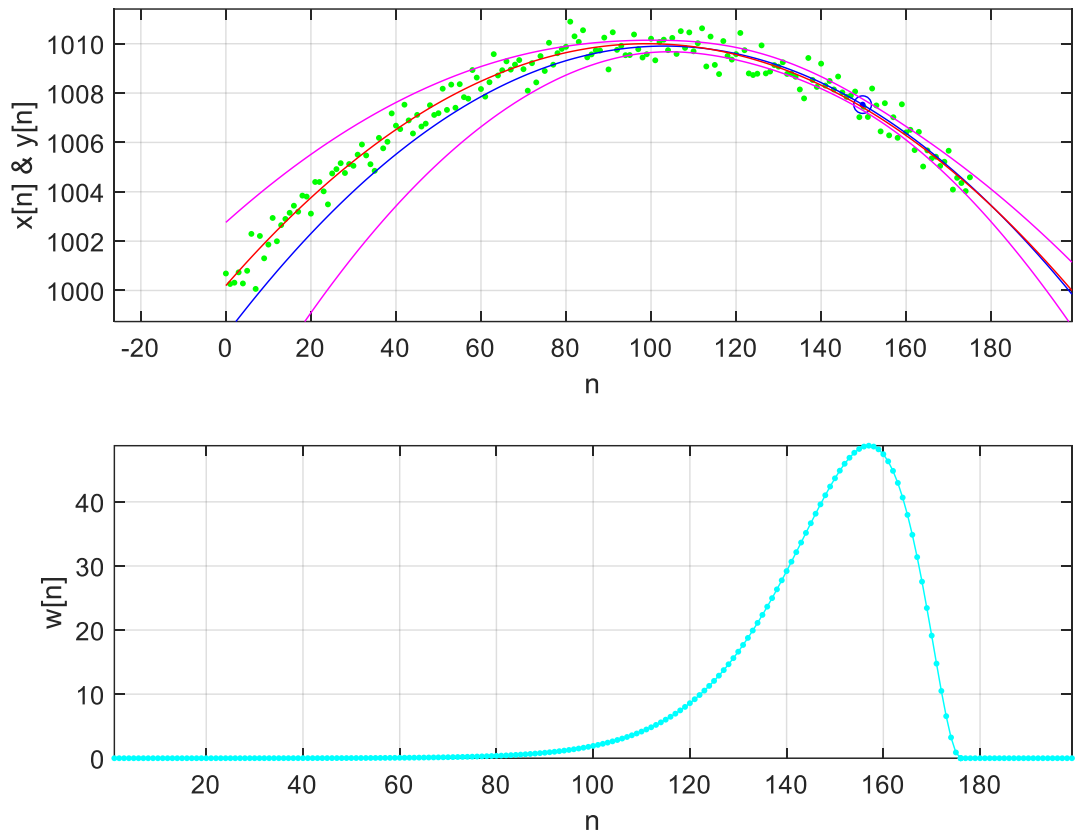

*Figure 16. Constant-acceleration tracking scenario. Low-gain state estimator with $K_X = 3$, $\kappa = 2$ & $p = 0.9$. Regression weight on lower subplot; weighted least-squares polynomial fit in upper subplot, with true target position (red), position estimate (blue) 3-sigma limits (magenta) and noisy measurements (green).*

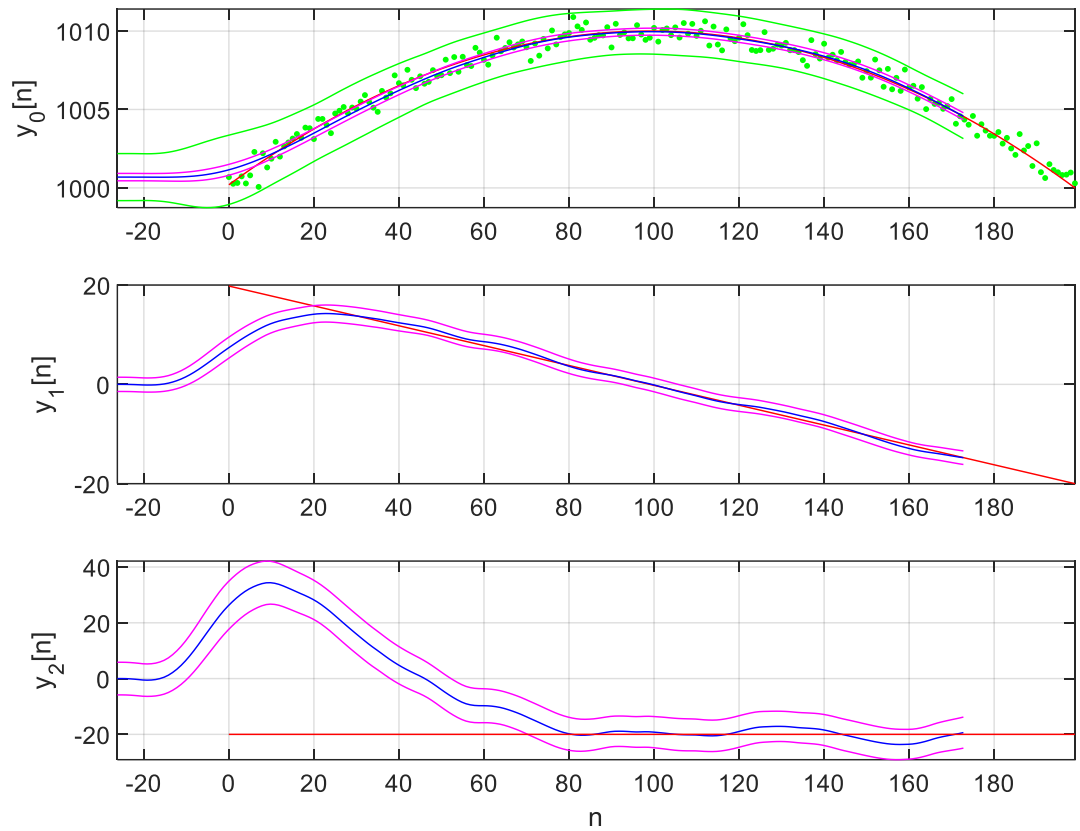

*Figure 17. Constant-acceleration tracking scenario. Output of low-gain state estimator with $K_t = 3$, $K_X = 3$, $\kappa = 2$ & $p = 0.9$. Position (with measurements in green), velocity and acceleration estimates and 3-sigma limits (blue and magenta lines) shown in the upper, middle and lower subplots, respectively.*

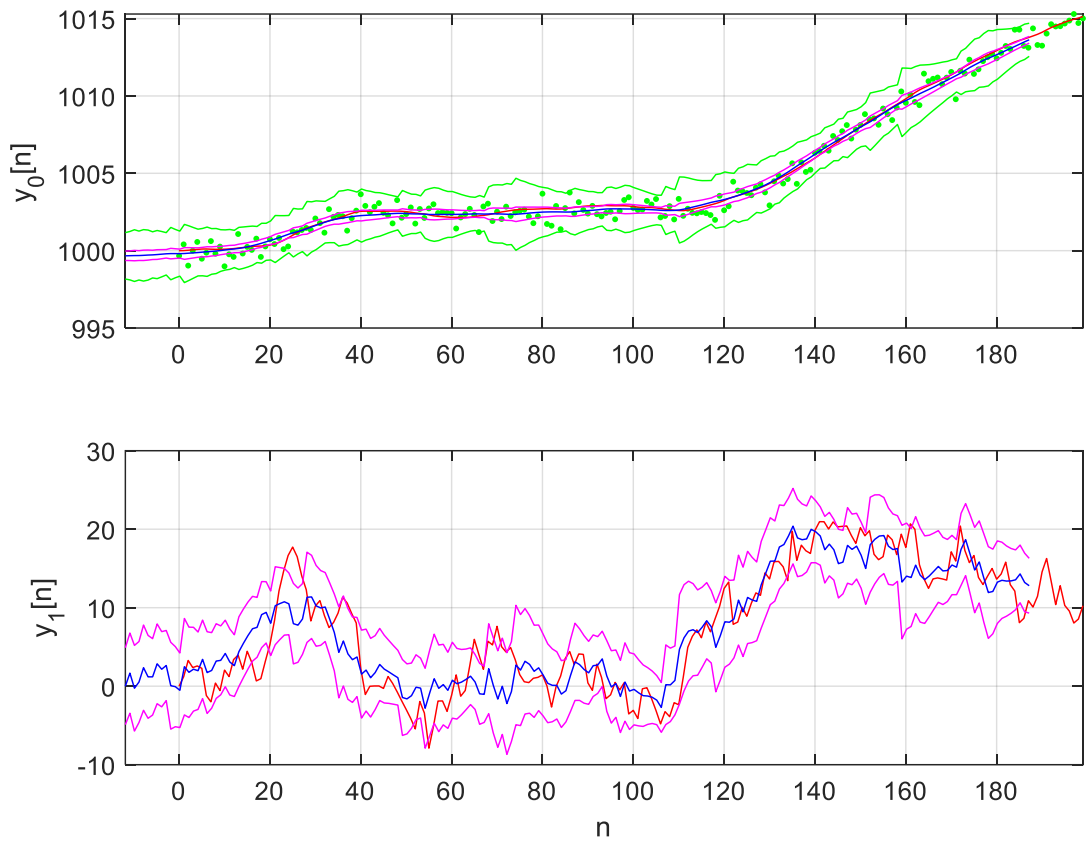

*Figure 18. Random-acceleration tracking scenario. Output of high-gain state estimator with $K_t = 2$, $K_X = 2$, $\kappa = 0$ & $p = 0.85$. Position (with measurements in green), velocity, and acceleration, estimates with 3-sigma limits (blue and magenta lines) shown in the upper, middle, and lower, subplots, respectively.*

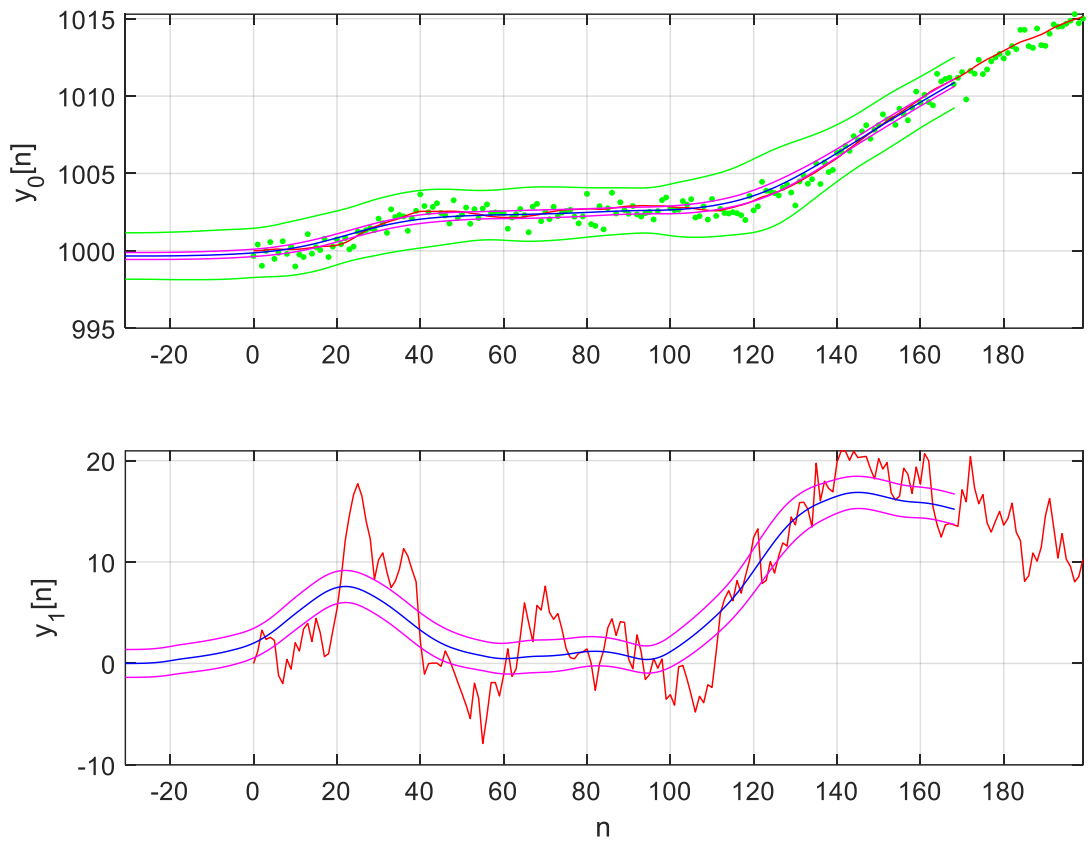

Figure 19. Random-acceleration tracking scenario. Output of low-gain state estimator with $K_t = 2$, $K_X = 2$, $\kappa = 3$ & $p = 0.85$. Position (with measurements in green), velocity, and acceleration, estimates with 3-sigma limits (blue and magenta lines) shown in the upper, middle, and lower, subplots, respectively.

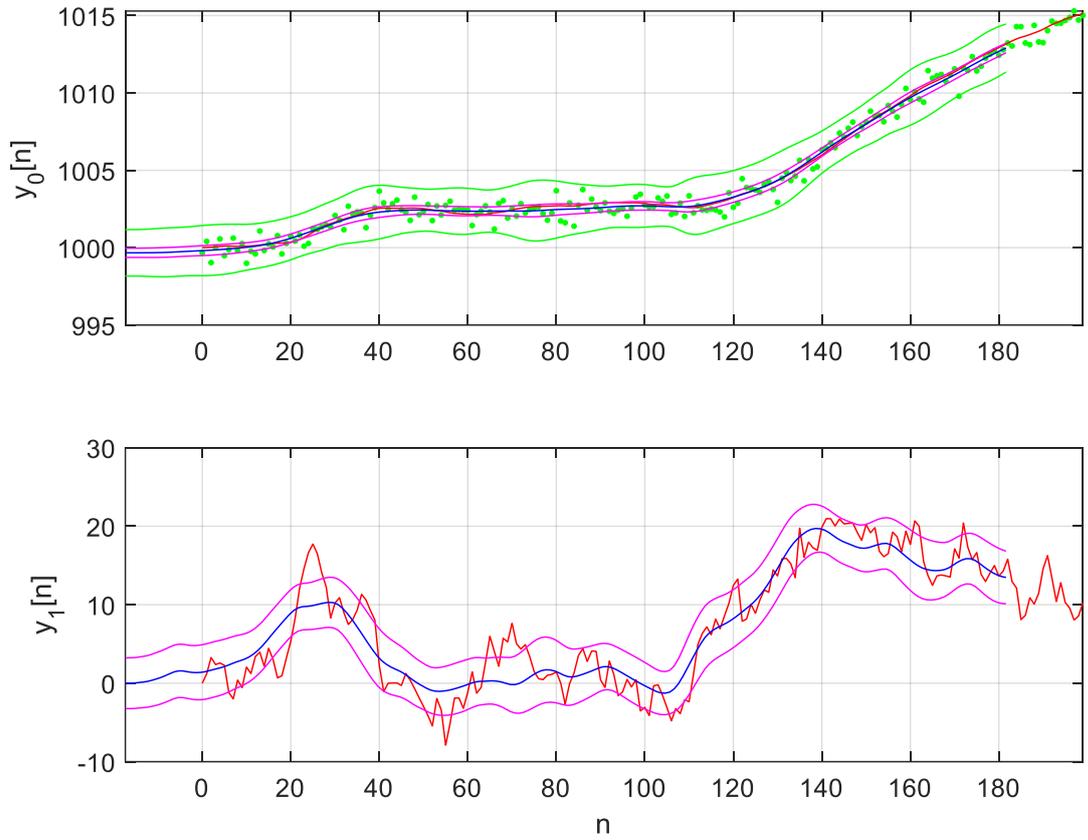

*Figure 20. Random-acceleration tracking scenario. Output of medium-gain state estimator with $K_t = 2$, $K_X = 2$, $\kappa = 3$ & $p = 0.75$. Position (with measurements in green), velocity, and acceleration, estimates with 3-sigma limits (blue and magenta lines) shown in the upper, middle, and lower, subplots, respectively.*

## Edge detector

Non-causal band-pass filters (e.g. low-order derivative filters) are commonly used to detect edges in two-dimensional images. In this use case we are interested in detecting the leading and trailing edge of a pulse in noisy time-series data, using causal band-pass filters, to detect its presence and determine its duration, for instance.

In this application example, a synthetic pulse is generated by convolving a rectangular pulse (100 samples long) with a Gaussian pulse (with a standard deviation of 8 samples) to smear the pulse edges. The resulting pulse is scaled to have a maximum amplitude of 20 and is superimposed on a constant background with an amplitude of 100. Gaussian measurement error is added with a standard deviation of 0.5 and the sampling period is 0.01 seconds. Regression filters with $K_X = 2$ (see Figure 21) and $K_X = 3$ (for a wider bandwidth, see Figure 23), are applied. Both filters use $K_t = 2$ with $\kappa = 2$ & $p = 0.8$.

The derivative estimate of the regression filter is used to generate a test statistic for each new sample. The dimensionless test statistic is defined as follows:

$Z = \hat{X}^{(1)}/\hat{\sigma}_1$ where (64a)

$\hat{X}^{(1)}$ is the estimate of the first derivative, see (46), and (64b)

$\hat{\sigma}_1^2$ is the error variance estimate of the first derivative estimate, see (49)-(52). (64c)

The null hypothesis is that the first derivative of the signal is zero. The null hypothesis is rejected, and an edge detection is declared, when $Z$ is large. This occurs when the polynomial fit is good, i.e. when the weighted sum of regression residuals is small and when the estimate of the first derivative is far from zero. A leading pulse edge is declared when $Z > \lambda_Z$; a trailing pulse edge is declared when $Z < -\lambda_Z$, where $\lambda_Z$ is an arbitrary threshold that is selected to yield an acceptable probability of detection and probability of false alarm. Logic to determine the stationary points of $Z[n]$, i.e. local maxima or minima, of threshold exceedances, so that only one detection is declared per crossing, is not considered here.

The test statistic generated using the estimator with $K_X = 2$ and $K_X = 3$ are shown in Figure 22 and Figure 24, respectively. The wider bandwidth of the $K_X = 3$ estimator allows the pulse edges to be followed with less bias error thus reduced variance, which results in a test statistic with taller and sharper peaks, that are resolved better, which should improve pulse edge detection, pulse edge localization, and pulse duration estimates.

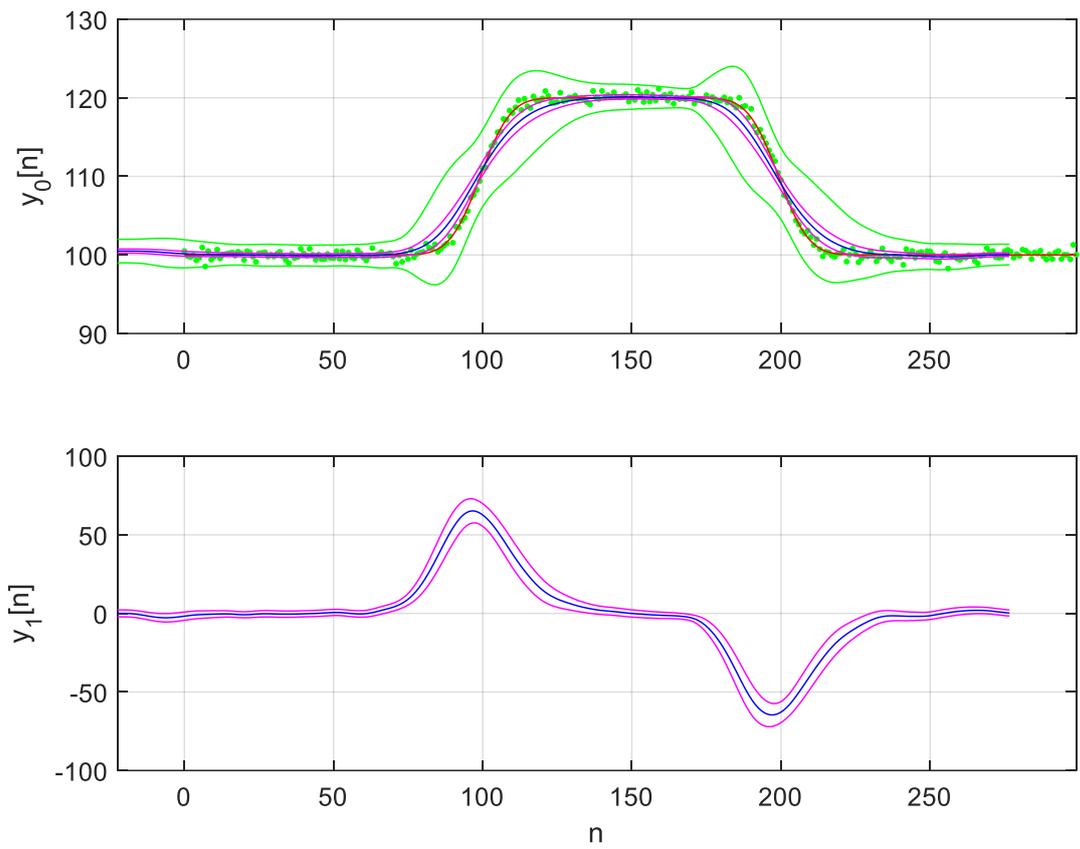

*Figure 21. Edge-detection use-case. Input measurements (green dots) and output estimates (blue and magenta lines). Output of smoother (upper subplot) and differentiator (lower subplot) for $K_X = 2$.*

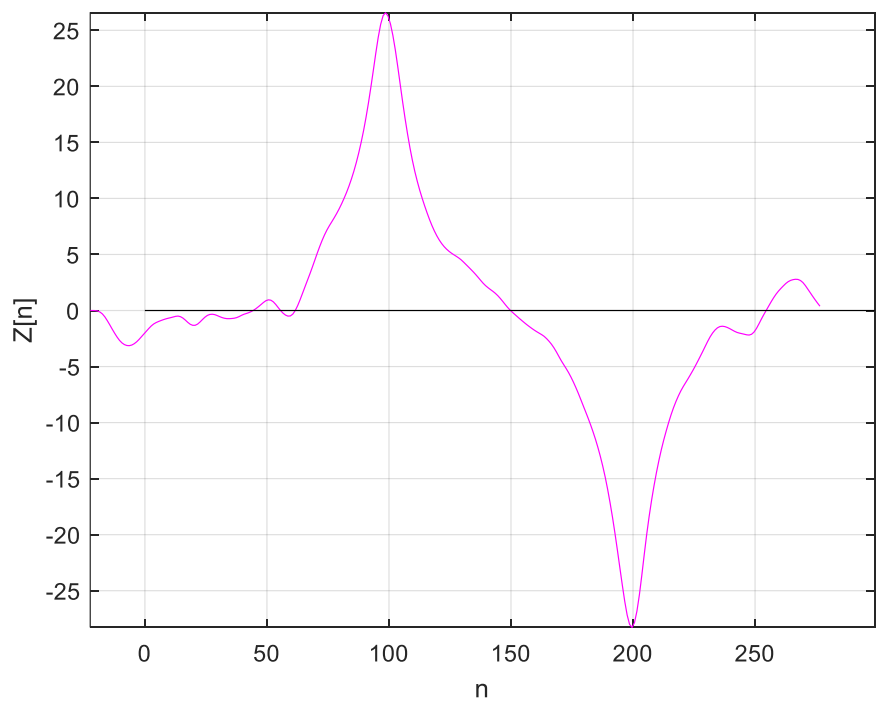

*Figure 22. Test statistic ($Z$) for pulse-edge detection, generated using the $K_X = 2$ estimator.*

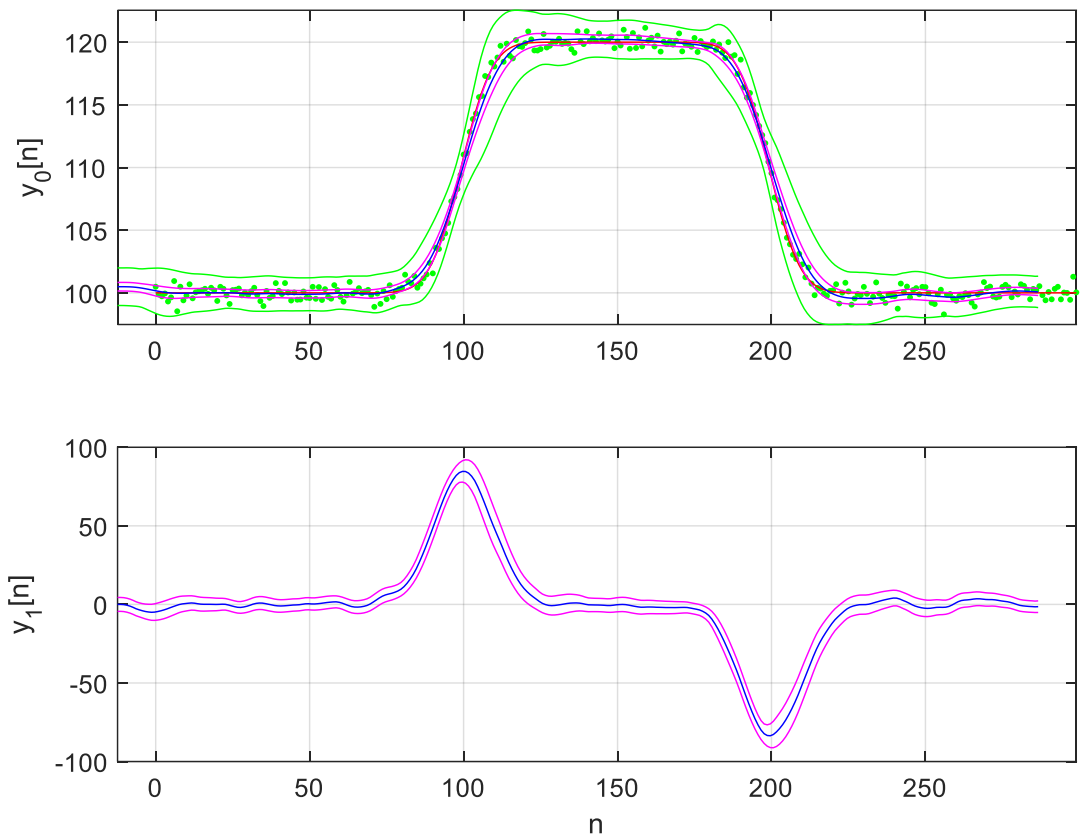

*Figure 23. Edge-detection use-case. Input measurements (green dots) and output estimates (blue and magenta lines). Output of smoother (upper subplot) and differentiator (lower subplot) for $K_X = 3$.*

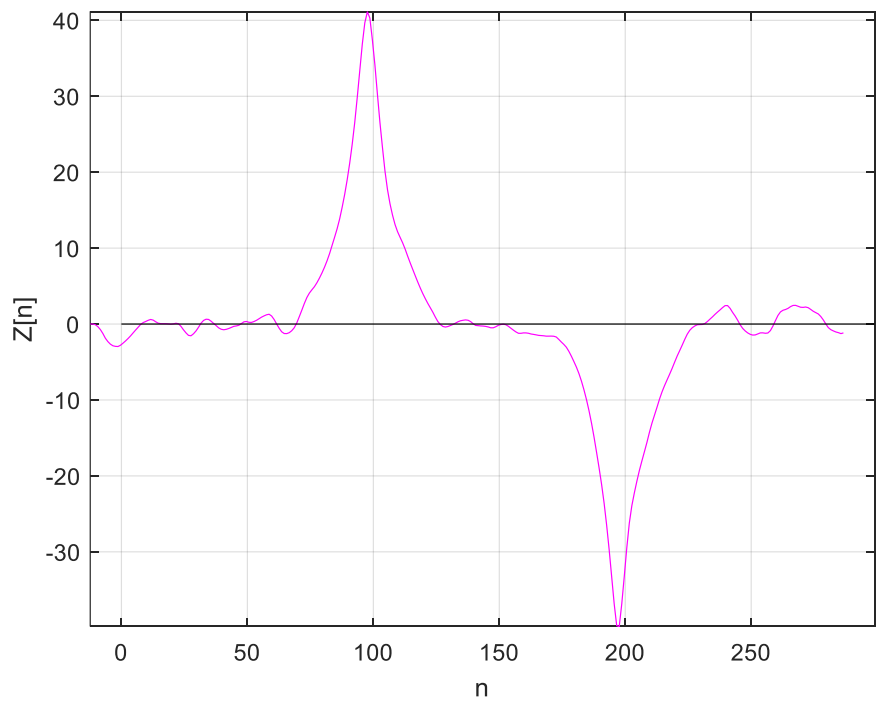

*Figure 24. Test statistic ($Z$) for pulse-edge detection, generated using the $K_X = 3$ estimator.*

## Peak detector

The detection and localization of significant local maxima (i.e. peaks) in noisy time-series data, that are superimposed on a fluctuating local background and interspersed with other insignificant local maxima, is another common problem in sensor signal processing, e.g. in biometric monitors.

In this example a 'target' peak and a 'clutter' spike were synthesised using Gaussians with an amplitude of 10 and standard deviations of 12 and 2 (samples) respectively. Random Gaussian noise with a standard deviation of 0.5 and low frequency 'interference' and were also added. The interference was generated using a polynomial with roots at $n = 0$, $n = (N-1)/2$ & $n = N - 1$, where $N$ is the number of samples in the synthetic time series ($N = 400$ & $T_s = 0.01$ seconds). The interference is scaled to have a maximum amplitude of 20 and a minimum amplitude of 10.

Target peaks cannot be extracted by simply applying and amplitude threshold to the time-series data, because this also would detect clutter and interference. Instead, the curvature of the waveform is estimated by fitting a quadratic polynomial then evaluating its second derivative. The scale of the regression weight is approximately 'matched' to the target peaks, i.e. the scale over which the peak is approximately quadratic (near its maximum). A regression filter with $K_t = 3$, $K_X = 3$, $\kappa = 2$ & $p = 0.8$ (for $\sigma_w = 7.762$ samples) was used for this purpose. The output of this filter is shown in Figure 25.

As in the edge detector use-case, a dimensionless test statistic is used to test the null 'curvature-is-zero' hypothesis. The test statistic in this peak-detector application is defined as follows:

$Z = -\hat{X}^{(2)}/\hat{\sigma}_2$ where (65a)

$\hat{X}^{(2)}$ is the estimate of the second derivative, see (46) and (65b)

$\hat{\sigma}_2^2$ is the error variance estimate of the second derivative estimate, see (49)-(52). (65c)

A peak is declared on local maxima that exceed a specified detection threshold, i.e. where $Z > \lambda_Z$.

The test statistic $Z[n]$ is plotted in Figure 26. Using $\lambda_Z = 10$ is sufficient to detect the target peak and suppress detections on the clutter spike and the background interference. The undulating background does not trigger the detector because the extent of the regression weight is insufficient to integrate the wide curve of the interference. Thus the interference power is insignificant relative to the noise power (see lower subplot of Figure 25). The clutter does not trigger the detector because a quadratic curve is a poor fit for the narrow spike over a much wider window. The sudden growth of the three-sigma limits indeed indicates that the quality of the fit around the clutter peak is poor (see upper subplot of Figure 25) and inference (e.g. detection declarations) should be suspended while this is the case. The target peak does trigger the detector because the quality of the quadratic fit is good over the scale of the regression weight and the negative curvature is significant, relative to the error estimate.

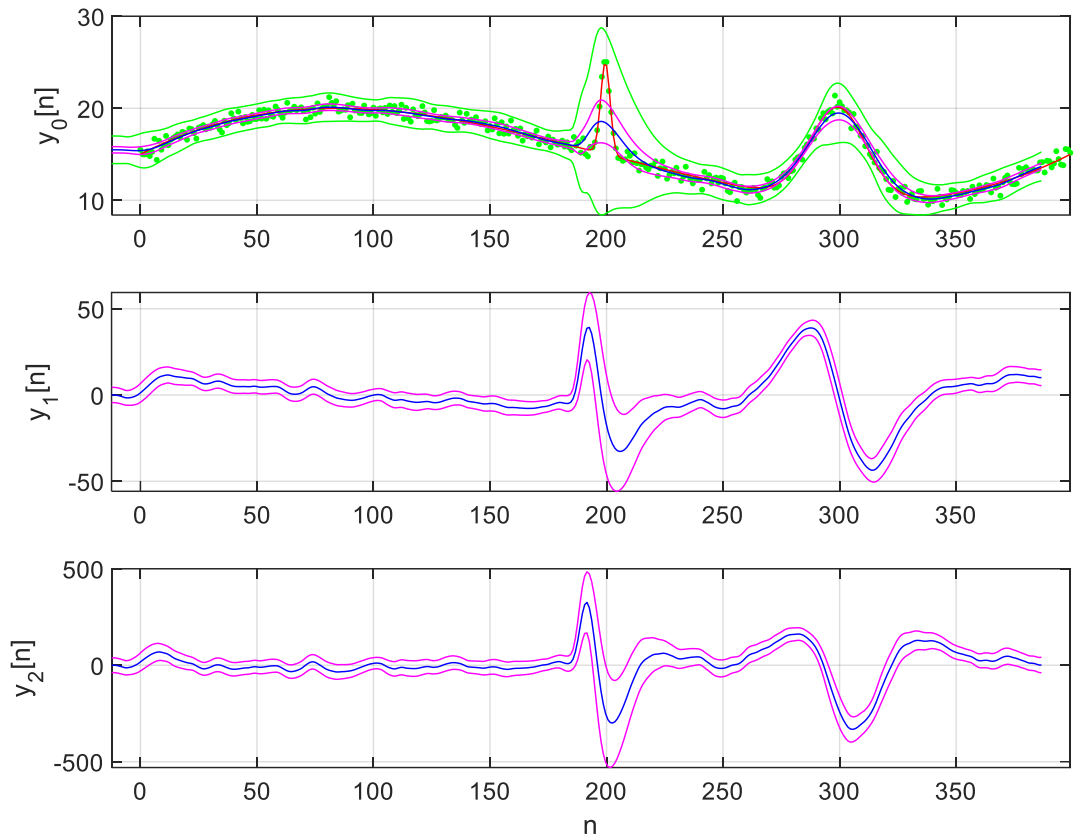

*Figure 25. Peak-detection use-case. Input measurements (green dots) and output estimates (blue and magenta lines). Output of smoother, first derivative, and second derivative are shown (top to bottom) for a regression filter with $K_t = 3$, $K_x = 3$, $\kappa = 2$ & $p = 0.8$. Clutter spike is at $n \cong 200$, target peak is at $n \cong 300$, both are on undulating low-frequency background interference.*

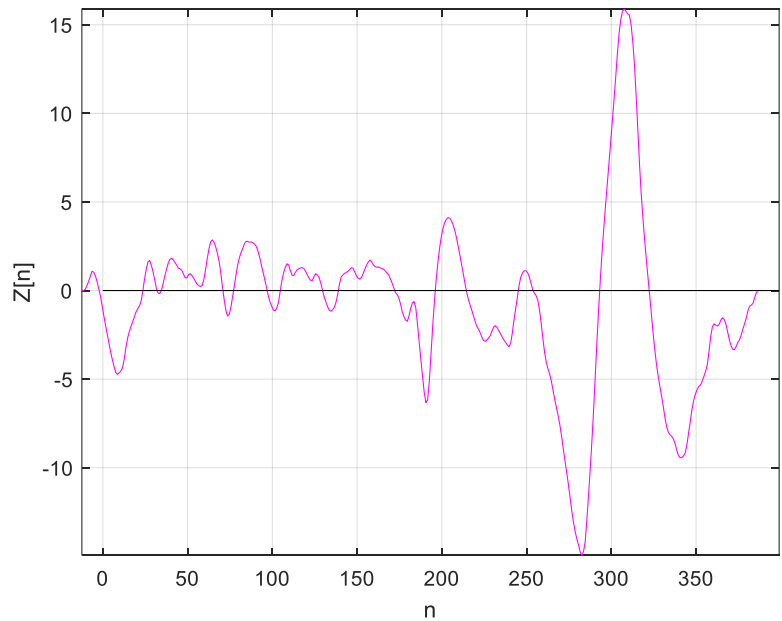

*Figure 26. Test statistic (Z) for peak detection.*

## Change detector

In this use case the aim is to detect significant changes in a trend, where a 'trend' is defined using a polynomial of specified degree. Such an algorithm may be used to detect pulses in electronic support, detect manoeuvres in a surveillance radar, or trigger orders in trading algorithms. The simulation parameters used in the previous pulse-detection use-case are again used here; however, the pulse waveform is now superimposed on a linear background, with an amplitude equal to $100 + n/10$, instead of a constant background of $100$.

The trend is defined using a polynomial of first degree, i.e. a model with $K_X = 2$, with unspecified $\alpha$ or $\beta$ coefficients, e.g. a constant velocity target. Two regression filters with different weights are applied to the data stream and their outputs are combined to form the test statistic. The weight applied in filter A emphasizes newer data whereas the weight applied in filter B emphasizes older data. At every $n$th sample, filters A and B estimate $X$ at the common time of $t = T_s(n - q)$, with $T_s = 0.01$ seconds, and a 'break' in the trend is declared when the estimates formed using newer and older data do not 'agree' with each other. The following dimensionless test statistic is used as a measure of 'disagreement':

$$Z = \frac{\hat{A} - \hat{B}}{\sqrt{\hat{a}^2 + \hat{b}^2}} \text{ where} \tag{66a}$$

$\hat{A}$ & $\hat{B}$ are the mean outputs of filters A & B for the $k_t = 0$ filter, i.e. $\hat{X}^{(0)}$, see (46) and  (66b)

$\hat{a}^2$ & $\hat{b}^2$ are the variance outputs of filters A & B for the $k_t = 0$ filter, i.e. $\hat{\sigma}_0^2$, see (49).  (66c)

A break up or down is declared on local maxima or minima that are above or below a specified detection threshold, i.e. where $Z > \lambda_Z$ or $Z < -\lambda_Z$; alternatively, $Z^2$ (a signal-to-noise ratio) may be used when the direction of the break is unimportant.

The parameters of filters A & B are shown in Table 10. Filter A uses a weight with $\kappa = 0$ to emphasise more recent data whereas filter B uses $\kappa = 3$ to emphasise less recent data. Both use the same smoothing parameter $p = 0.8$ (see Figure 4) for a realization with a common feedback loop for the recursive generation of the weighted sums, i.e. the **G** & **H** matrices defined in (38a) & (41b). The outputs of filters A & B are formed using different **C** vectors, which applies the required magnitude scaling the phase shift to the weighted sums. Reducing the delay of filter B, i.e. shifting its output forward in time, so that it matches the delay of filter A, increases the noise gain (i.e. the VRF) and the bandwidth (i.e. $f_c$). For this common delay, the VRF of filter B is greater than that of filter A but the $f_c$ of filter B is less than that of filter A. The higher VRF for a lower $f_c$ is due to a magnitude 'bulge' on the edge of the low-frequency passband of filter B; however, over medium to high frequencies the magnitude of filter B is less than filter A, except at $f = 1/2$ where filter A has a null (see Figure 27). Both filters have the required magnitude flatness and phase-linearity in the very low-frequency passband; however, the phase lag of filter B is much greater than that of filter A, at higher frequencies, due to the lagged response of the Erlang window with a non-zero shape parameter.

The outputs of filters A & B for the simulated input are shown in Figure 28 & Figure 29. These outputs are plotted on the same axes in the upper subplot of Figure 30; the numerator and denominator of (66a) are plotted in the middle subplot and the resulting test statistic is plotted in the lower subplot. Using $\lambda_Z = 3$ would be sufficient to detect the beginning and end of the leading and trailing pulse edges in this simulation.

Table 10. Parameters of regression filters used in change detection use-case.

| $K_X$ | 2 | 2 | 2 |
|---|---|---|---|
| $\kappa$ | 0 | 3 | 3 |
| $p$ | 0.8 | 0.8 | 0.8 |
| $q$ | 8.50 | 22.41 | 8.50 |
| VRF[0,0] | 0.0556 | 0.0305 | 0.0725 |
| $f_c$ | 0.0420 | 0.0256 | 0.0352 |

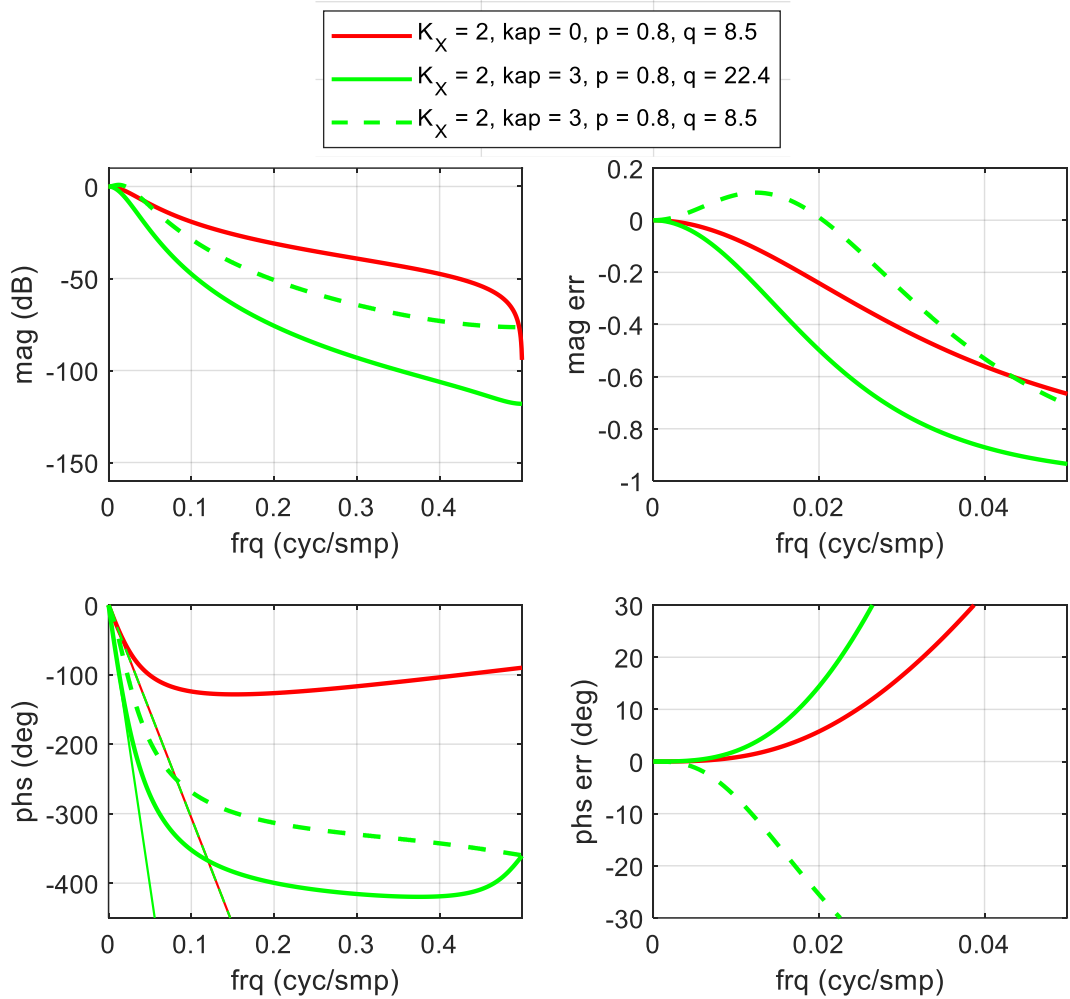

Figure 27. Frequency response of regression filters used in change detection use-case. Filter A with optimal delay (sold red), filter B with optimal delay (solid green, not used in simulations) and filter B with a delay matched to filter A (dashed green).

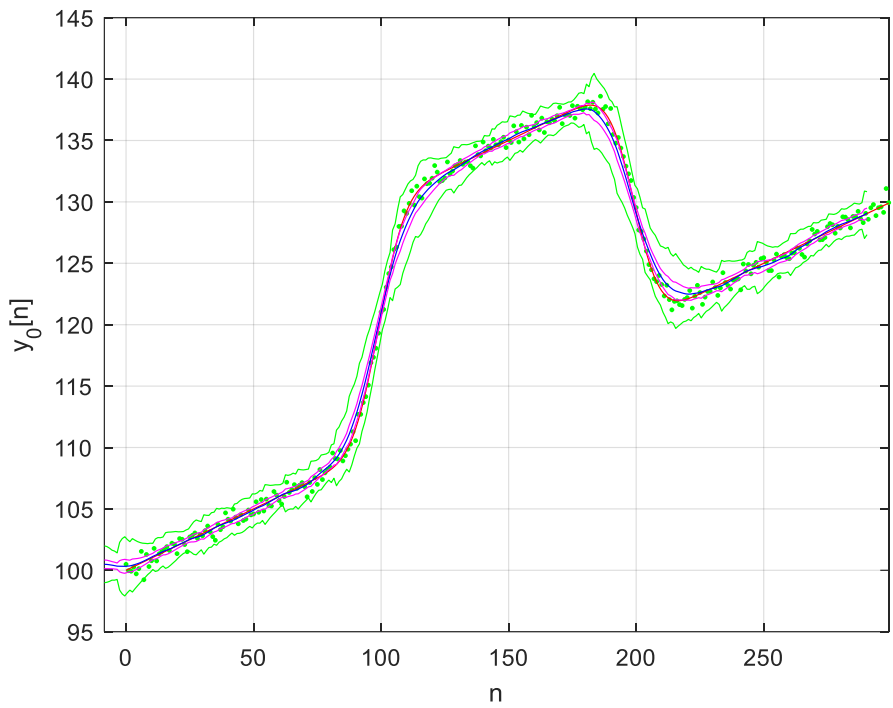

*Figure 28. Output of filter A, i.e. $\hat{A}$. Smoothed estimate is formed with newer data emphasised using $\kappa = 0$. Fitted polynomial is evaluated using $q = 8.5$.*

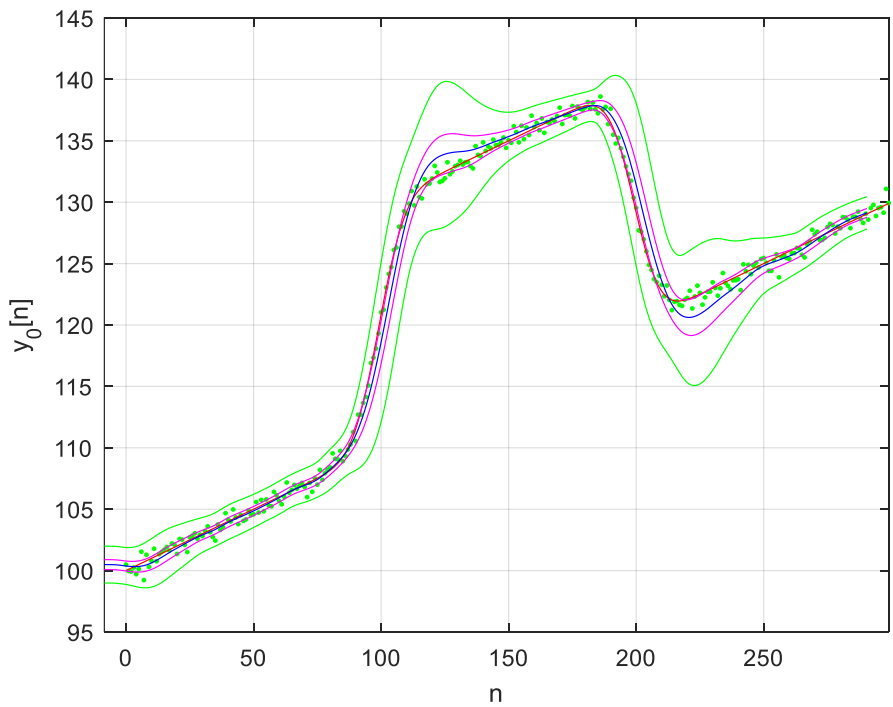

*Figure 29. Output of filter B, i.e. $\hat{B}$. Smoothed estimate is formed with older data emphasised using $\kappa = 3$. Fitted polynomial is evaluated using $q = 8.5$.*

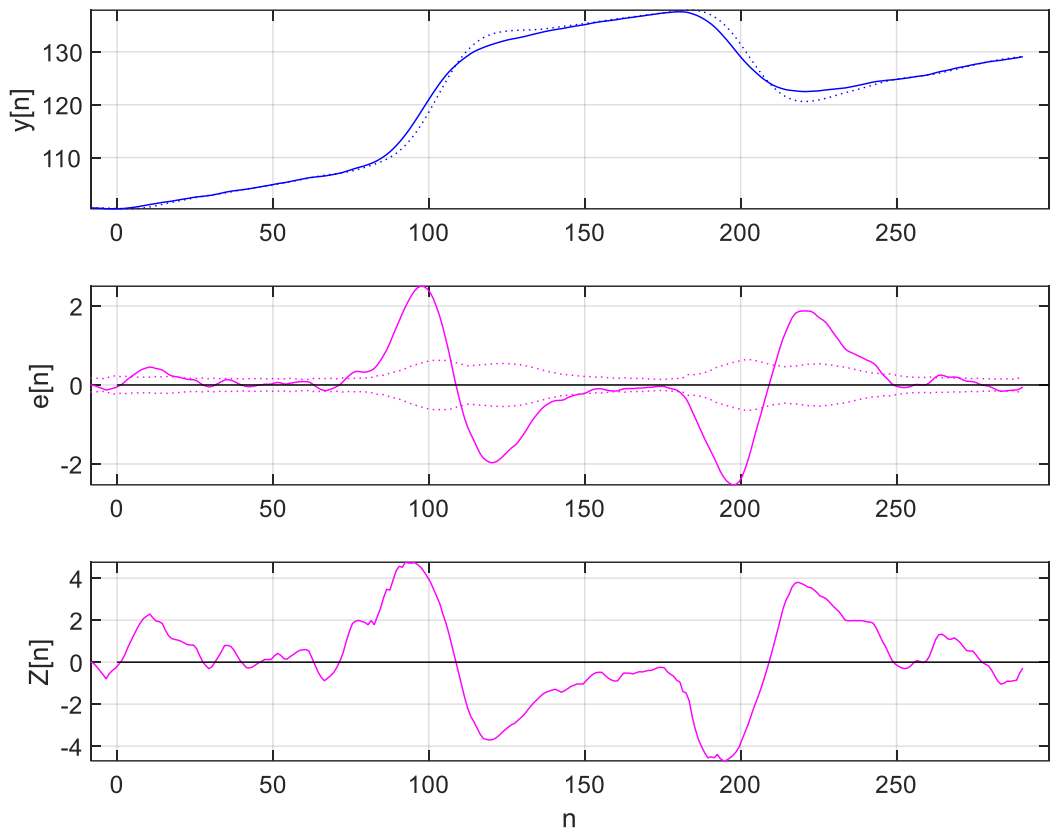

*Figure 30. Illustration of change-detection algorithm. Upper subplot: filter A estimate (solid blue) and filter B estimate (dotted blue), i.e. $\hat{A}$ & $\hat{B}$. Middle subplot: estimate difference, i.e. $\hat{A} - \hat{B}$ (solid magenta) and the square root of the variance sum, i.e. $\sqrt{\hat{a}^2 + \hat{b}^2}$ (dotted magenta). Lower subplot: the test statistic, i.e. $(\hat{A} - \hat{B})/\sqrt{\hat{a}^2 + \hat{b}^2}$ (magenta).*

## Closing remarks

It is easier to interpret and appreciate the significance of a given moment in time when it is considered in context. The past and future are of equal importance. Reflection upon the past comes at no cost, although we eventually reach a point where the past is no longer relevant. Hindsight is a wonderful thing, although there may be a cost of inaction while we wait, albeit briefly, for the future to unfold.

A flexible regression weight manages these considerations and shapes the magnitude and phase response (e.g. the bandwidth and lag) of the resulting filters with an infinite impulse response (IIR). Relative to a weight with a simple exponential decay (with $\kappa = 0$) that is usually used in 'discounted' recursive regression filters, more general Erlang weights (with $\kappa \geq 0$) offer improved noise attenuation, without sacrificing bandwidth, in applications where longer lags (as set using the delay parameter $q$) are acceptable.

The use cases in Section 5 were selected to illustrate the potential and flexibility of simple low-pass (smoothers) and band-pass (derivative) filters in a diverse range of signal-processing problems. The use cases also highlight the additional leverage obtained by computing and using the variance for an indication of estimate quality. The tuning guidelines in Section 4 were intended to elucidate the way in which the principles of weighted polynomial regression may be used to shape the impulse and frequency response of the realized filters. The mathematical foundations of derivative state estimation and weighted polynomial regression were outlined in Sections 3 & 2 and low-complexity feedback structures for their recursive implementation in software, firmware, or hardware, were described in Section 1.

## Acknowledgements

This work was sponsored in part by the Defence Science and Technology Group.